\documentclass[preprint]{aastex}

\newcommand{\ha}{H$\alpha$}
\newcommand{\alphairac}{$\alpha_{IRAC}$}
\newcommand{\accunits}{$M_{\sun} yr^{-1}$}

\begin{document}
\title{Evidence for Early Circumstellar Disk Evolution in NGC2068/71}
\author{Flaherty, K.M. \altaffilmark{1}, Muzerolle, J. \altaffilmark{1}}
\email{kflaherty@as.arizona.edu}

\altaffiltext{1}{Steward Observatory, University of Arizona, Tucson, AZ 85721}

\begin{abstract}
We study the disk and accretion properties of young stars in the NGC 2068 and NGC 2071 clusters. Using low-resolution optical spectra, we define a membership sample and determine an age for the region of $\sim$2 Myr. Using high-resolution spectra of the \ha\ line we study the accretion activity of these likely members and also examine the disk properties of the likely members using IRAC and MIPS mid-infrared photometry. A substantial fraction (79\%) of the 67 members have an infrared excess while all of the stars with significant infrared excess show evidence for active accretion. We find three populations of evolved disks (IRAC-weak, MIPS-weak and transition disks) all of which show decreased accretion activity in addition to the evidence for evolution in the dust disk.

\end{abstract}  

\section{Introduction}
Circumstellar disks are a common, if not universal, part of the star formation process. The dust and gas within them provide the building blocks for planetary systems , hence their study is of critical importance for understanding planet formation processes and timescales. A considerable amount of work has been done in the past 20 years to inventory circumstellar disks in star forming regions and young stellar clusters, constraining their properties and lifetimes as a function of stellar age, mass, and environment. The advent of the Spitzer Space Telescope, with its unprecedented mid-infrared sensitivity, has opened up the field in this regard and is now providing the most complete census of disks to date.

In young clusters the disk fraction is a strong function of age, decreasing from $\sim80\%$ for clusters at 1 Myr down to almost no stars with disks for clusters older than 10 Myr \citep{hll01,her07}. However, large variations in the disk properties exist at any age. Observations of the infrared spectral energy distribution (SED), which traces dust emission from the disk within a few AU of the star, show a wide range of excess for single clusters \citep{mea05,sa06,dah07,cur07,gor07, win07, bal07} as well as a decrease in the median excess with age \citep{sa06,her07,la06}. Submillimeter observations \citep{bec91,miy93,and07} and spectra of the 10\micron\ silicate feature \citep{bou01,kes06,sar06} also show a large variation in the dusts properties between circumstellar disks. The exact nature of this evolution, and why some stars lose their disks before others, has not been completely solved. Strong dependences on the steller mass have been found, particularly at ages $>$2-3 Myr, where less massive stars tend to hold onto their disks longer \citep[e.g.][]{la06,car06}. The role of environment is less clear; while some disks are clearly destroyed more quickly in strong photoionizing environment, there is some evidence that the overall disk frequency changes little \citep{bal07}.

Much of the recent Spitzer studies concerning disk evolution have focused on somewhat older regions with ages $>$2 Myr \citep{la06,mea05,sa06,dah07}. In order to investigate earlier disk evolution, we present here an analysis of the stellar population of a young embedded cluster associated with the NGC 2068 and 2071 reflection nebulae (hereafter treated as one region NGC 2068/71). As a site of ongoing star-formation with hundreds of young stellar objects, this region provide a unique environment to study early disk evolution. As part of the much larger Orion B molecular cloud, they are located at a distance of approximately 400 pc \citep{ant82}. The large number of sources with infrared excess indicate it is still a very young cluster. Studies of dense gas \citep{lad91} and dust continuum \citep{joh01} reveal a large number of dense molecular clumps and cores, consistent with the youthful nature of this region.

We have analyzed optical spectra along with optical, near and mid-infrared photometry to identify members of the NGC 2068/71 cluster and study the stellar and circumstellar properties of these members. Section ~\ref{data} describes the data we have obtained, section~\ref{results} describes our results derived from these data, and section~\ref{discussion} includes a discussion of the implications of these results.

\section{Data\label{data}}

Low-resolution optical spectra were taken with the Hectospec instrument on the MMT. Hectospec is a fiber-fed multi-object spectrograph that uses 300 fibers to cover a 1$^{\circ}$ field of view \citep{fab05}. These spectra have a resolution of R=1200 and cover the wavelength range 3700-9150\AA. They were obtained on Nov 5,2004 and Oct 23,2005 with an exposure time of 3x900 sec. Our sample selection methodology will be discussed below.

High-resolution optical spectra were taken with the Hectochelle instrument on the MMT. Hectochelle uses 240 of Hectospec's 300 fibers to cover a similar field of view at much higher resolution \citep{sze98}. Spectra were centered on the \ha\ line at 6563$\AA$ covering a wavelength range of 6460-6648\AA\ with a resolution of R=32,000. Two fields, each with an exposure time of 3x1200 sec, were observed on Dec 3,2004 and Nov 14,2005. Five stars (\#177, 326, 458, 618 and 1116) were included in both high-resolution observing runs in order to examine the variability in the \ha\ profile of these objects.

Near-infrared $JHK_s$ photometry was taken from the Two-Micron All Sky Survey (2MASS). Mid-infrared photometry was obtained using the IRAC and MIPS cameras aboard the {\it Spitzer} Space telescope. The IRAC observations were taken as part of GTO program 29, IRAC Survey of the Orion Molecular Clouds. IRAC images were taken in High-Dynamic Range (HDR) mode with exposures of 10.4 and 0.4 seconds \citep{faz04}. Individual BCD frames (pipeline version S9.5.0) were mosaicked using a custom IDL program creating separate mosaics from the short exposures and the long exposures. Source selection on these mosaics was completed using PhotVis v1.08 \citep{gut04}. Photometry was derived from individual BCD frames using the highest signal to noise data that did not have any saturated pixels (Megeath et al. in prep). The photometry used an aperture of 2.4'', and a sky annulus with inner and outer radii of 2.4'' and 7.2''. Magnitude zero points of 21.93, 21.26, 19.08, 19.44, including the aperture correction, were used for these data.

The MIPS observations were taken in scan mode using medium scan rate and half-array offsets, with a total areal coverage of about 40'x45' common to the three arrays.  The total effective exposure times are 80, 40, and 8 seconds at 24, 70, and 160 microns.  Since only a few Class II sources were detected at 70 microns and none at 160, we have omitted those data in this study. The images were reduced using the MIPS instrument team Data Analysis Tool, which includes basic calibration as well as mosaicking of individual exposures \citep{gor05}.  The final mosaic was created with half-pixel subsampling.  We then performed photometry on individual point sources in the 24 micron mosaic using PSF fitting with an emipirical PSF constructed from bright isolated point sources in the field.  A PSF fit radius of 5.6" and sky annulus of 15-22.5" were used.  An aperture correction of 1.73, as estimated from the STiny Tim PSF model \citep[e.g.][]{eng07}, was applied to the raw photometry.  Fluxes were converted to absolute units using a conversion factor of 150 DN/s/mJy.

Optical photometry was taken from the Sloan Digital Sky Survey (SDSS) \citep{yor00}. The Orion field was observed as part of an early calibration run through the Galactic plane in the fall of 1998 and the fall of 1999 \citep{fin04}. SDSS photometry includes measurements at {\it ugriz}, whose effective wavelengths are 3540, 4760, 6290, 7690, 9250\AA. In total we have photometry extending from 0.35-24\micron\ over a field roughly $0.8^\circ\times1.2^\circ$ centered on the NGC 2068 and 2071 nebulae. The MIPS field is slightly smaller, which only affects a few of our selected cluster members. Table~\ref{member_phot} lists the photometry for the cluster members and Figure~\ref{poss_image} shows our cluster members overlayed on an optical image of the cluster.

\section{Results\label{results}}

\subsection{Membership Selection}
\begin{figure}
\epsscale{.6}
\plotone{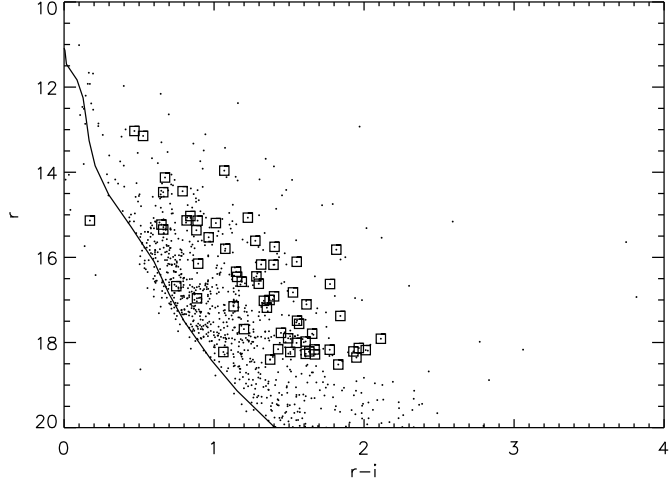}
\caption{SDSS color-magnitude diagram for those stars with optical photometry (dots) and those stars we classify as likely members (squares). Most of the likely members lie above the main sequence locus \citep{s00}, as expected for these young stars.\label{sdss_cmd}}
\end{figure}

Targets for the spectroscopic survey were chosen based on their position on the r vs. r-i color magnitude plot above the expected position of the main sequence at the distance of NGC 2068/71, as seen in Figure~\ref{sdss_cmd}. We restricted ourselves to stars with r magnitude between 13 and 18.5 to exclude stars that would saturate and those that are too faint to be easily observed. The faint limit corresponds to an unextincted 2 Myr old M6 star at 400pc. Of these stars many are foreground/background stars unrelated to the cluster. The cluster members were identified from the presence of  Lithium (6707 \AA) absorption and \ha\ emission in the low-resolution spectra. Since \ha\ emission can be seen in older stars, such as dMe stars, we use strong Li absorption as the primary indicator of youth, while \ha\ and other emission lines are used as a secondary indicator. Lithium absorption is an accurate indicator of youth in low-mass stars, but is more ambiguous for G type stars and earlier; with smaller convective envelopes these stars do not burn the Li in their atmospheres as quickly and Li absorption can be observed in older stars. Along with our flux limit, this focuses our survey to low mass stars of spectral type K to mid M. We do not attempt to characterize the high mass or brown dwarf population of young stars within NGC 2068/71.

\begin{figure}
\epsscale{.6}
\plotone{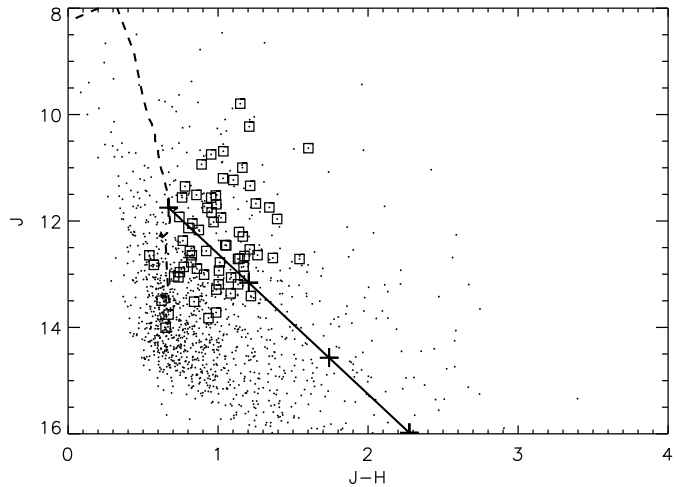}
\caption{NIR color-magnitude diagram for all sources observed toward NGC 2068/71 (dots) and all members (squares). The dashed line is the 2 Myr isochrone from \citet{s00} at the distance of NGC 2068/71. The solid line is a reddening vector extending from a 2 Myr M0 star at the distance of NGC 2068/71, with every 5 $A_V$ marked with an plus sign.\label{nir_cmd}}
\end{figure}

\begin{figure}
\epsscale{.6}
\plotone{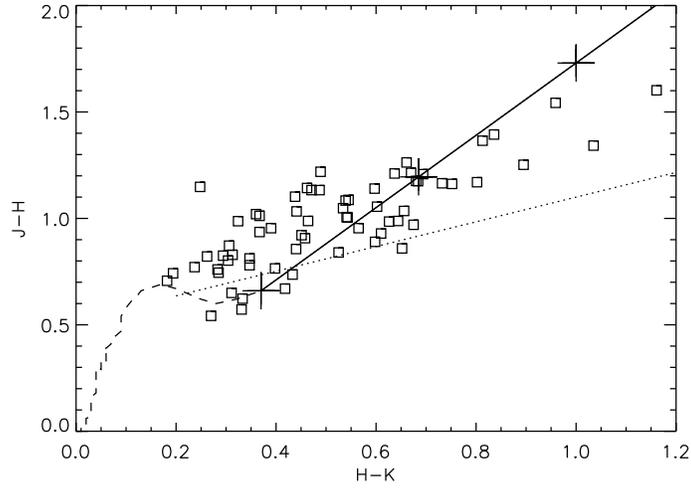}
\caption{NIR color-color diagram of the clusters members. Dashed line shows the main sequence locus from \citet{kh95} and the dotted line is the CTTS locus from \citet{m97}. The solid line is the reddening vector extending from an M6 star with every $A_V=5$ marked with a plus sign. The one star with anamolously blue colors is object 458 which may be contaminated by scattered light in the near-infrared.\label{nir_ccd}}
\end{figure}

Our survey is also limited by extinction. Since we define our members based on optical spectra, we cannot survey the deeply embedded population. Figure~\ref{nir_cmd} shows a J vs. J-H color-magnitude diagram, similar to Figure~\ref{sdss_cmd}, with a reddening vector extending from a 3 Myr old M0 star at 400 pc. Our sample is limited to $A_V<5$ although many sources further along the reddening vector have excess emission at 3.6-24\micron\ consistent with circumstellar disks and hence are probably embedded cluster members. A J-H vs. H-K color-color diagram for the cluster members is shown in Figure~\ref{nir_ccd}. The cluster members without a near-infrared excess fall within $A_V=5$ of the main sequence locus due to extinction limit imposed by the use of optical spectra.

 Table~\ref{lowres} lists the stars identified as likely cluster members along with their spectral types and Li and \ha\ equivalent widths. Of the 453 stars observed spectroscopically, 69 are likely members. They are a mix of classical and weak T-Tauri stars (CTTS,WTTS respectively) and range in spectral type from G6-M6. Two are continuum stars that do not show any significant features that could be used to derive a spectral type. They are added as members, but are not included in the rest of our analysis. CTTS and WTTS are separated based on their \ha\ equivalent width, using the \ha\ EW versus spectral type boundary from \citet{wb03}. Resolved \ha\ profiles are needed to best discriminant between accretors and non-accretors, as will be discussed later.

\subsection{HR Diagram}

To construct an HR diagram and estimate an age for NGC 2068/71, the luminosity and effective temperature are derived for each star. The effective temperature is derived from the strength of multiple absorption bands and lines within the low-resolution spectra, based on the prescription of \citet{sa04} with a few modifications. In the prescription of \citet{sa04} the spectra are split into broad categories of early ($<$K1), medium (K1-M0) and late (M0-M6) spectral types, and separate bands/lines are used within each broad category to derive an effective temperature. Indices were formed for these bands based on the ratio of flux in the feature to flux in the continuum. We did not use the TiO2 (4975\AA) band for the medium spectral types or the TiO1 (4775\AA) band for late spectral types because they produced $T_{eff}$ that were inconsistent with the other bands. The spurious temperature from the TiO1 band could be caused by low signal to noise in the later spectral types or veiling from accretion shocks, while the spurious temperature from the TiO2 band could be caused by H$\beta$ emission in the continuum interval next to the feature. The TiO3 (6185\AA) and TiO4 (7140\AA) bands are best used for stars with TiO1 index $>1.2$ and TiO2 index $>1.26$ indicative of later spectral types, but we used them whenever possible because of their tight correlation between band strength and $T_{eff}$. When multiple bands were used to determine the $T_{eff}$ of the star the spread in $T_{eff}$ was used to estimate the uncertainty in spectral type. Spectral types of the members were visually confirmed by comparing to the standard spectra of \citet{as95}. The spectral types of M stars were also double-checked with the CaII and CaH spectral bands of \citet{as95}, as well as the VO bands of \citet{wi05}. Effective temperatures were transformed to spectral types using the conversion of \citet{kh95}. 

Veiling can sytematically effect our spectral typing since it will reduce the band strengths, resulting in an earlier spectral type. We can estimate the effect of veiling for those star where measurements of the veiling exist (see section~\ref{sec_accrete}). Since veiling is roughly constant redward of 5000\AA\ \citep{bas90} our measurements of the veiling at 6500\AA\ should be equal to the veiling at the bands that are used to measure spectral type. For these stars the effect of veiling is on the order of the spectral type uncertainty. Overall we do not account for veiling when deriving spectral type, but we do not expect it to substantially change our results.

The luminosity of each star was calculated from its optical photometry, transformed from SDSS ugriz magnitudes to Johnsons-Cousins UBVRI magnitudes \citep{jor06}. The extinction was derived using the observed R-I color along with the intrinsic R-I color from \citet{kh95} and the extinction law of \citet{rl85}. We chose the extinction law of \citet{rl85} because it forms the basis for more detailed extinction law studies \citep{ccm89}. It also is comparable to recent determinations of the extinction law in the 2MASS passbands \citep{ind05} allowing us to use it when deriving reddening from R-I colors and from JHK colors (discussed below). The luminosity was then determined from the dereddened I band using bolometric corrections from \citet{kh95} and a distance of 400 pc. Two objects, \#984 and 458, have no optical photometry and no optical luminosity is calculated. Table~\ref{member_params} lists the luminosities, effective temperatures and radii, derived assuming L=$4\pi R^2\sigma T_{eff}^4$. The typical uncertainty in effective temperature is 5\%, while the typical uncertainty in luminosity, which includes uncertainty in $T_{eff}$ and photometric uncertainty, is 10\%.

The luminosity can also be derived from the 2MASS J band magnitude. To determine the amount of extinction for each star, the observed J-H color was dereddened to the intrinsic J-H color, taken from \citet{kh95}, for the WTTS while the CTTS were dereddened to the CTTS locus from \citet{m97}. Bolometric corrections from \citet{kh95} were used  and a distance of 400 pc was assumed. There is no systematic difference between the extinction or luminosity derived from the optical colors or the near-infrared colors. For three stars where dereddening based on R-I colors resulted in negative extinction, i.e. the observed colors are bluer than the intrinsic colors, the extinction was instead derived from the near-infrared. Two of these stars have late spectral types (M5-M6) for which a different choice of intrinsic colors (e.g. \citet{luh03}) would result in positive extinction. For these stars the extinction is close to zero and a change in the intrinsic colors will have a significant effect on the reddening derived. Object 1171 is much brighter in the optical than in the near-infrared, possibily due to variability, and the extinction and luminosity are both derived from the near-infrared. For objects \#984 and 458 the extinction and luminosity were derived from the near-infrared photometry because of the lack of optical photometry. Object 458 appears extended in the 2MASS images giving it anamolous colors as seen in figure~\ref{nir_ccd}, and its extinction is assumed to be zero even though it is likely to be much higher. Since the extinction is most likely non-zero the luminosity we derive is a lower limit. For the rest of the stars we use the optically derived luminosities for our analysis. 

\begin{figure}
\epsscale{.6}
\plotone{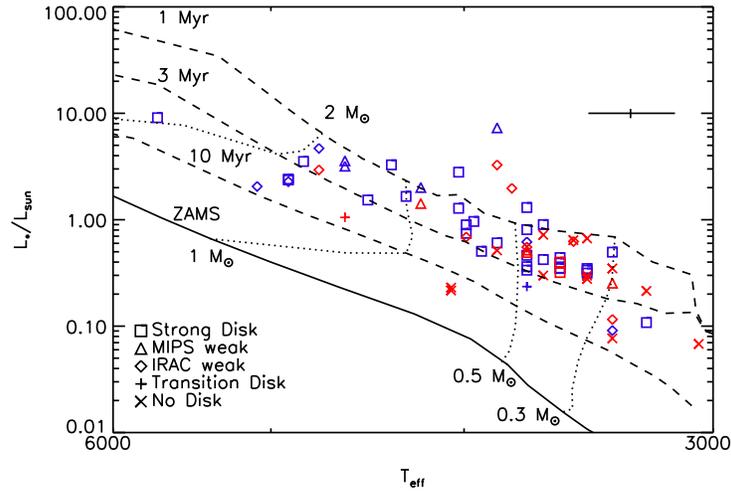}
\caption{HR diagram for the likely members of NGC 2068/71. Different symbols represent different types of disks. Blue symbols are CTTS and red symbols are WTTS. See text for definition of strong/IRAC-weak/MIPS-weak/no disk as well as description of derivation of luminosity and effective temperature. Dotted lines show mass tracks for a 0.3, 0.5, 1, 2 $M_{\sun}$ star. A typical error bar is shown in the upper right corner. \label{hr}}
\end{figure}

Figure~\ref{hr} shows the HR diagram for NGC 2068/71 along with theoretical isochrones from \citet{s00}. The members are spread from above the 1 Myr isochrone to below the 10 Myr isochrone, a spread beyond our errors in luminosity and temperature and typically seen in HR diagrams of similar regions \citep{hil97}. There appears to be no significant age difference between the CTTS or WTTS. The median age for the cluster is 2 Myr, with the majority of the stars lying between the 1 and 3 Myr isochrones. Using optical spectroscopy limits our study to the stars that are not heavily embedded within the molecular cloud. Our measured age of 2 Myr may be an overestimate of the true mean age if the substantial population of embedded stars is systematically younger. Without any information about their luminosities or $T_{eff}$ we use an age of 2 Myr for the rest of our analysis.

When studying the age and masses of the stars in NGC 2068/71 we have chosen to use the theoretical isochrones of \citet{s00}. Other isochrones, such as those by \citep{bar98} and \citep{dan97} would result in a different estimate for the age of the cluster and the masses of indivual stars. The D'Antona \& Mazzitelli tracks tend to produce younger ages, with most of the NGC 2068/71 members lying above the 1 Myr isochrone. The Baraffe et al. models agree more accurately with dynamical mass estimates of binaries, with large scatter, \citep{hil04} although the models do not extend to our most massive members. The Siess et al. tracks measure sytematically low masses, similar to the D'Antona \& Mazzitelli tracks, although the tracks do follow the locus of cluster members in the HR diagram. We do not account for these systematic effects but only note that the age of the cluster and the masses of individual stars is model dependent.

Contamination from the nearby older, 4-6 Myr, OB1b association might affect our estimate of the age of NGC 2068/71. Without proper motion it is difficult to distinguish members of NGC 2068/71 from members of OB1b that have strayed into our field of view. We can estimate the contamination by considering the radial distribution of Orion OB1b members from \citet{bri05}. Based on the size of our NGC 2068/71 field and distance from the center of the OB1b association we expect roughly one star from the OB1b association to fall within our field. This will not substantially affect our estimate of the age of the cluster, or our study of the disk population.

The mass of each star was derived based on its position in the HR diagram by interpolating along a theoretical isochrone between the nearest model masses using effective temperature. This was done for the isochrone above and below each member, and the two mass estimates were averaged. Isochrones at 1, 2, 3, 5, 10, and 20 Myr were used while stars above the 1 Myr isochrone were assumed to lie on the 1 Myr isochrone and stars below the 20 Myr isochrone were assumed to lie on the 20 Myr isochrone. There is little difference in the mass estimate when using the two isochrones because many of these stars are young enough to still lie on vertical convective tracks. For those higher mass stars that lie on older isochrones there is more uncertainty in the mass estimate because this position in the HR diagram corresponds to an evolution along constant luminosity rather than constant $T_{eff}$. For these stars, typically with M$>1 M_{\sun}$, we interpolated along each isochrone using luminosity instead of $T_{eff}$. Since most of our stars are K to mid-M this only applied to the handful of stars with the earliest spectral types. The typical uncertainty in the mass estimates is 10\%, based on the uncertainty in effective temperature, ignoring systematic uncertainties in the theoretical models. Masses are listed in Table~\ref{member_params}.

\subsection{Radial and Rotational Velocity}
\begin{figure}
\epsscale{.6}
\plotone{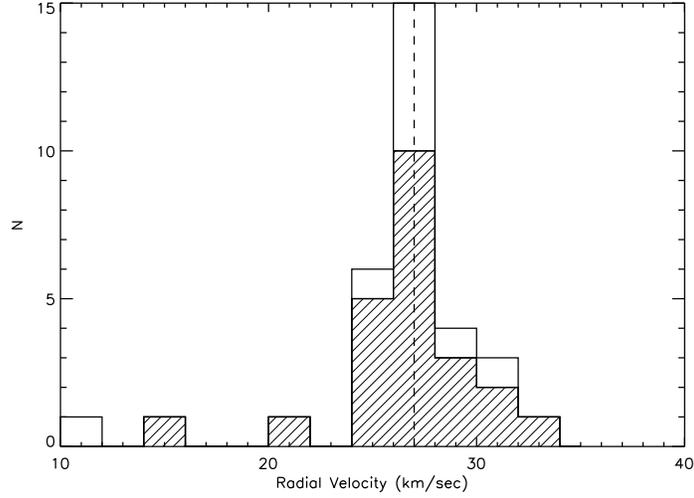}
\caption{Radial velocity histogram for the likely members. Dashed line shows the median velocity of likely members at 27.1 km/sec. Hashed histogram shows the CTTS in the cluster, which make up most of the likely cluster members around this median velocity.\label{vr}}
\end{figure}

Radial velocities were derived for those stars with high-resolution spectra using the cross-correlation program {\it fxcor} in IRAF, along with 3 late G, early K type standard stars. \citet{ha86} describe in detail the methodology used to derive radial and rotational velocities. Our results are given in Table~\ref{highres}. Two objects (\#584,373) do not have any spectral lines above the noise that can be used to measured an accurate radial velocity, and no radial velocity is recorded. The distribution of radial velocities for all members is shown in Figure~\ref{vr}, along with just the CTTS, which show a similar distribution. A gaussian fit to the radial velocities of the members has a center at 27.1 km/sec and a standard deviation of 1.3 km/sec, which is similar to the uncertainties in the radial velocities. A similar spread is measured in the radial velocities of Taurus and ONC members \citep{ha86,sa05}. The radial motion of the members is consistent with the motion of the surrounding molecular cloud \citep{lad91,wil05}. Given the youth of the members, they have not had much time to travel outside of the cloud in which they were born. If we assume the velocity spread represents the difference between stellar and cloud velocities, it would take 5 Myr to travel 5 pc, roughly the size of the molecular cloud \citep{wil05}.

\begin{figure}
\epsscale{.6}
\plotone{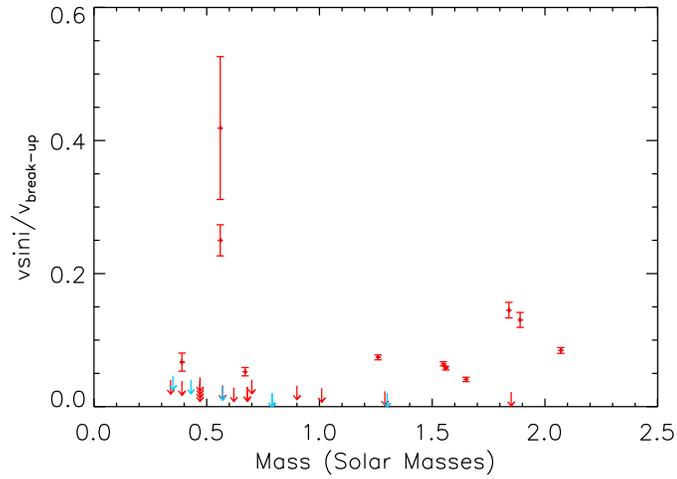}
\caption{Rotational velocity as a fraction of the break-up velocity ($v_{breakup}=(GM/R)^{1/2}$) versus mass. Upper limits are marked with arrows. Red symbols indicate stars with disks based on the presence of infrared excess and blue symbols represent stars without disks.\label{vsini}}
\end{figure}

Rotational velocities were derived from the width of the cross-correlation function determined by the program {\it fxcor}. The relationship between the width of the cross-correlation function and the rotational velocity is determined by artificially broadening spectral standards, which all have a rotational velocity below the resolution limit of 8 km/sec. A quadratic function was fit to the width versus vsini, and the average of these fits for the three spectral standards was used to convert the width of the correlation function for the object spectra to vsini values. The \ha\ line was excluded when measuring the rotational velocity since it is not broadened by rotation alone. Rotational velocities, listed in Table~\ref{highres}, are plotted as a function of break-up velocity ($v_{breakup}=(GM/R)^{1/2}$) in Figure~\ref{vsini}. Most of the members have vsini below the resolution limit of 8 km/sec and almost all have a rotational velocity which is a small fraction of the break-up velocity. The two stars with the highest fractional rotational velocity have noisy continua, which could lead to a spuriously high rotational velocity. The rotation rates are consistent with the small vsini seen in other regions of similar ages \citep{sa05,sa06,cla00}.

\subsection{Accretion Activity}\label{sec_accrete}

The \ha\ line is a sensitive measure of accretion activity. For our sample the equivalent width (EW) of the \ha\ line, measured from the low-resolution spectra and listed in Table~\ref{lowres}, was used to differentiate between actively accreting stars (CTTS) and stars that are not actively accreting (WTTS). We have used the EW boundary as a function of spectral type between CTTS and WTTS defined by \citet{wb03}. Self-absorption generally does not affect the EW. When the absorption is the strongest, the emission is also very strong because the accretion and mass loss that the two components trace are related to each other. With this in mind resolved \ha\ profiles are still the preferred accretion diagnostic.

\begin{figure}
\epsscale{.6}
\plotone{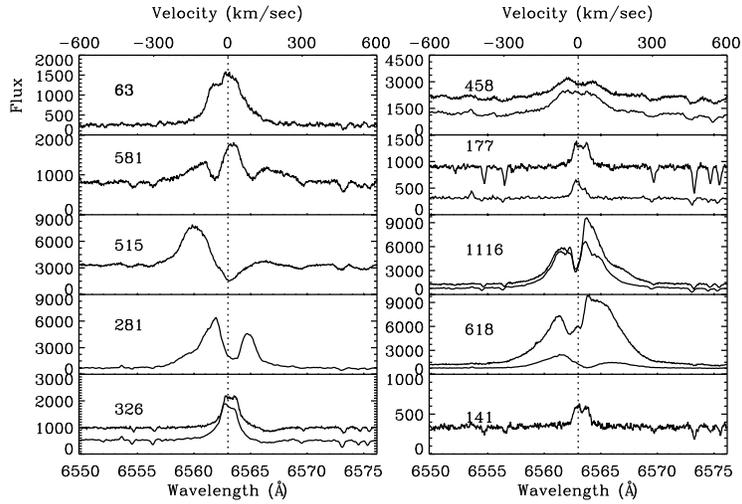}
\caption{Sample hectochelle spectra focusing on the \ha\ profiles. The stars include a strong disk (63), a MIPS-weak disk (581), an IRAC-weak disk (515), a transition disk (281), the five stars with repeat hectochelle spectra (326, 458, 177, 1116 and 618) and one star without a disk (141). Stars \#177, 141 demonstrate the narrow, symmetric profiles due to chromospheric emission while the rest show broad, asymmetric features due to accretion. Object ID numbers are included in each plot, along with a vertical line marking the line center, and the spectra of stars with multiple observations have been shifted vertically for easier comparison. \label{spec_ex}}
\end{figure}

High-resolution hectochelle spectra, examples of which are shown in Figure~\ref{spec_ex}, were available for 34/67 of our likely members, as well as our two continuum stars. For these stars the \ha\ profile was used to delineate active accretors from non-accretors instead of the \ha\ EW. A full width at 10\% maximum flux of 250 km/sec was used to separate accretors and non-accretors. Measuring the width at 10\% maximum flux rather than half maximum helps to avoid any self-absorption features in the profile while still measuring the profile above the noise of the continuum. \citet{wb03} use a boundary of 270 km/sec, while we choose 250 km/sec to consistently include veiled stars as actively accreting, see below and Figure~\ref{hafw}. Four exceptions (\#1262, 326, 994, 998 )out of the 35 members with high-resolution spectra show signs of accretion although their high-resolution \ha\ EW lies below the CTTS boundary. Object 1262 (\ha\ EW=8\AA\, FW 10\% =340 km/sec) has a very large absorption dip in the center of the \ha\ profile, reducing the flux in the line, and the measured EW, while object 326 (\ha EW=5.9\AA, FW 10\%=170 km/sec) has an inverse P Cygni profile. The presence of a strong absorption feature, such as the inverse P Cygni profiles in objects 326 and 515 in figure~\ref{spec_ex}, can lower the velocity width below our threshold of 250 km/sec. In these cases the presence of strong absorption features due to accretion flows, as well as an asymmetric profile, indicate that accretion is still ongoing. Object 984 has a very noisy high-resolution spectra, possibly due to its high reddening, $A_V=4.5$, making it difficult to evaluate its accretion status. Object 998 also has a noisy profile and the high-resolution spectra of these two stars was binned to increase the signal to noise. The width of the \ha\ profile is above our threshold for active accretion; however the accretion status of these stars is still questionable without a higher signal to noise \ha\ profile, or another sign of active accretion such as a U band excess. Unfortunately the red leak in the SDSS u-band prevents us from accuratly measuring the U-band excess in this filter. The \ha\ EWs measured from the high-resolution spectra, listed in Table~\ref{highres}, are consistent with the \ha\ EW measured from the low-resolution spectra.

Continuum veiling of the photospheric absorption features provides an additional measure of accretion activity, albeit one that is less sensitive than the \ha\ profile. Accretion shocks produce a featureless continuum that adds to the photosphere, reducing the strength of photospheric lines relative to the observed continuum. The change in these photospheric features can be measured by comparing a veiled spectra with an unveiled spectra of the same spectral type. The method of \citet{hart89} was used to calculate the veiling ($r=F_{excess}/F_{cont}$) over the wavelength range 6490-6501\AA. This wavelength range was chosen because it contains a number of strong stellar lines while other wavelength ranges in our high-resolution spectra did not consistently have strong lines useful for veiling and were ignored. We added a constant flux to the unveiled spectra until the strength of the photospheric features relative to the continuum matches that of the veiled spectra. This constant flux level relative to the continuum is designated r. The uncertainty in the veiling is approximately 0.15, and was estimated by comparing standards of similar spectral types to each other. Standards of spectral type K1.5,K2.5,K7,M0,M1 were used and veiled spectra were compared to standards within one spectral type. For the veiled stars that have a measurable rotational velocity, the standard was artificially broadened to match the rotational velocity of the veiled star. We found veiling measures ranging from 0.31 to 3.2, similar to those found near 6500\AA\ by \citet{har91}.

\begin{figure}
\epsscale{.6}
\plotone{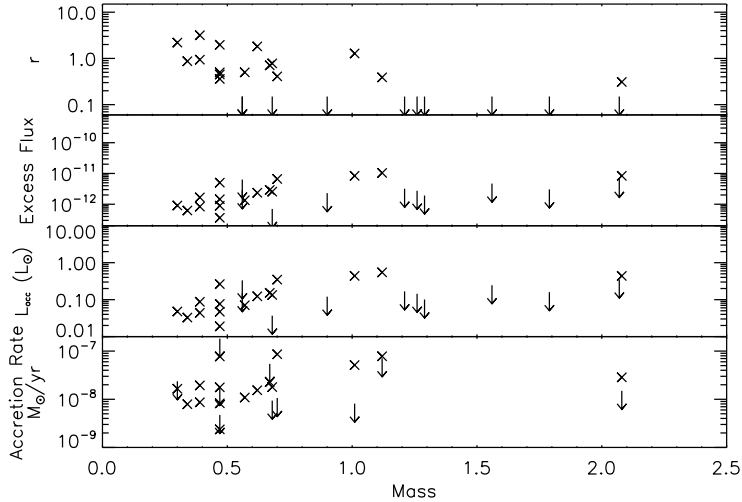}
\caption{Veiling, accretion luminosity and accretion rate as a function of mass. Upper limits are marked with arrows. See text for description of conversion from veiling to accretion rate.\label{accrete}}
\end{figure}

Accretion rates were estimated from the veiling measures using the procedure of \citet{wh04}. The conversion starts by using the R band magnitude as a measure of the continuum emission in this wavelength range, then from the definition of $r$ we determine the excess flux, converting to an accretion luminosity by assuming the total accretion flux was 11 times larger than the flux observed over the R band and then assuming the accretion infall comes from $3R_{*}$ to convert accretion luminosity to accretion rate. Correcting the observed accretion flux for the total accretion flux depends on the model of the accretion shock chosen. We use a correction factor of 11, a logarithmic average of various models, which is consistent with the analysis of \citet{wh04}. The radius $3R_{*}$ is typical for the truncation of the disk by the magnetosphere, where material from the disk flows onto the magnetic field lines \citep{ken96,wh04}. Instead of using theoretical R band magnitudes, we used observed R band magnitudes, correcting for the contribution from the veiling and extinction. This avoids the assumption that all of the stars have exactly the same age, and is not subject to the systematic uncertainties associated with theoretical isochrones. For those stars with no measurable veiling but signs of accretion, such as a wide and/or asymmetric \ha\ profile, an upper limit on the accretion rate was calculated assuming a veiling of r=0.15. Accretion rates are listed in Table~\ref{highres} and veiling, excess flux from accretion in the R band, accretion luminosity and accretion rate are plotted in Figure~\ref{accrete} as a function of mass. The $\dot{M}$ values are typical of the range in young regions such as Taurus \citep{gull98}. We do not cover a large enough range in mass to look for any statistically significant trends.

Five of our likely members, objects 326, 177, 458, 618 and 1116 have two high-resolution spectra taken a year apart. Object 618 has the largest variation with both a change in the shape and strength of the profile, even though the width of the profile does not significantly change, as seen in Figure~\ref{spec_ex}. The other stars do not vary significantly in the shape or strength of the \ha\ profile.  The 10\% FW vary by less than 40 km/sec for all of these stars, similar to other stars with similar \ha\ profiles \citep{joh95,jay06}.

Accretion is often associated with outflows and jets driving material away from the star \citep{kon00}. The NGC 2068/71 cluster has a number of strong molecular outflows \citep{whi81,sne82,bal82,fuk86} as well as Herbig-Haro objects \citep{zha99,rei99} throughout the cloud. None of these outflows appear to be associated with the pre-main sequence stars studied here, and may instead be related to the younger, less evolved stellar population of NGC 2068/71. In the optical spectra, a number of stars exhibit [OI] 6300\AA\ emission associated with outflows. These stars, marked in table~\ref{lowres}, all have an infrared excess and are actively accreting supporting the view that accretion and outflows are physically connected.

\subsection{Disk emission}

\begin{figure}
\epsscale{.6}
\plotone{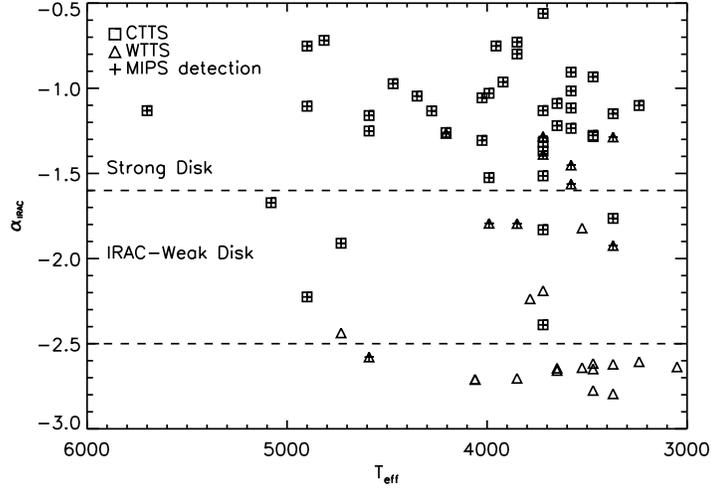}
\caption{Slope of IRAC SED versus effective temperature for the likely members. CTTS are marked with a square, WTTS are marked with a triangle. Dashed lines mark boundaries between stars with strong or MIPS-weak disks, stars with IRAC-weak disks and stars without disks.\label{irac_sed}}
\end{figure}

The shape of the infrared SED contains information about warm dust surrounding the young star and can be used to distinguish between stars with disks and stars without disks. \citet{la06} and \citet{her07} used the slope of the SED in the IRAC bands, \alphairac, to distinguish between stars with strong infrared excess coming from optically thick accretion disks, stars with a weaker excess indicative of evolved, or ``anemic'', disks, and stars with no measurable excess associated with a lack of disk material within 1 AU. We refer to the evolved objects as IRAC-weak because they may have strong excess at longer wavelengths indicative of optically thick outer disks. Figure~\ref{irac_sed} shows $\alpha_{IRAC}$ versus effective temperature for NGC 2068/71. As in \citet{la06} we find two distinct populations, one around \alphairac$=-1.0$ and a second around \alphairac$=-2.8$. We have not dereddened the SEDs since the low extinction ($A_v<5$) and relatively flat extinction law across the IRAC bands \citep{fla07} means that dereddening will have a small effect on the SED slopes. Previously a boundary of \alphairac=-1.8 has been used to distinguish evolved disks from strong disks \citep{la06,her07}. We use a boundary of \alphairac=-1.6 because of the stars near \alphairac=-1.8 that likely have some evolution in the structure of their disks. For our analysis stars with \alphairac$>-1.6$ were treated as having strong disks, stars with $-2.5<$\alphairac$<-1.6$ have IRAC-weak disks and stars without disks have \alphairac$<-2.5$.

\begin{figure}
\epsscale{.6}
\plotone{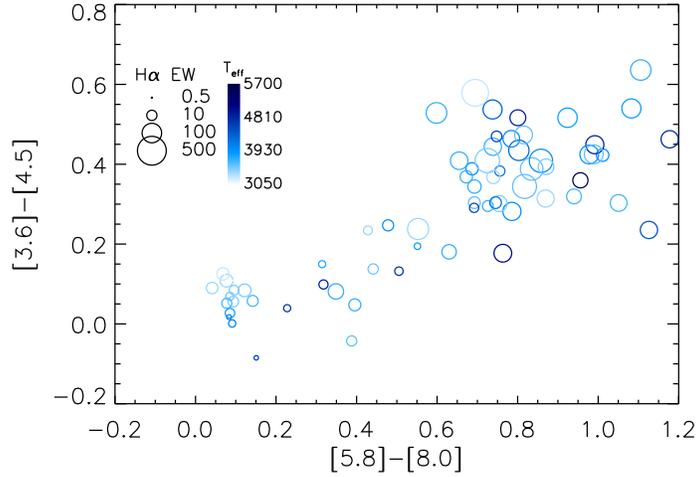}
\caption{IRAC color-color diagram for those stars in our spectroscopic sample. Size of the symbol indicates strength of \ha, as measured by its equivalent width, while the color of the symbol scales with its $T_{eff}$. The stars with the smallest infrared excess tend to have the smallest \ha\ EW, while the stars with the largest infrared excess in both colors have the largest \ha\ EW.\label{irac_ccd_haew}}
\end{figure}

\begin{figure}
\epsscale{.6}
\plotone{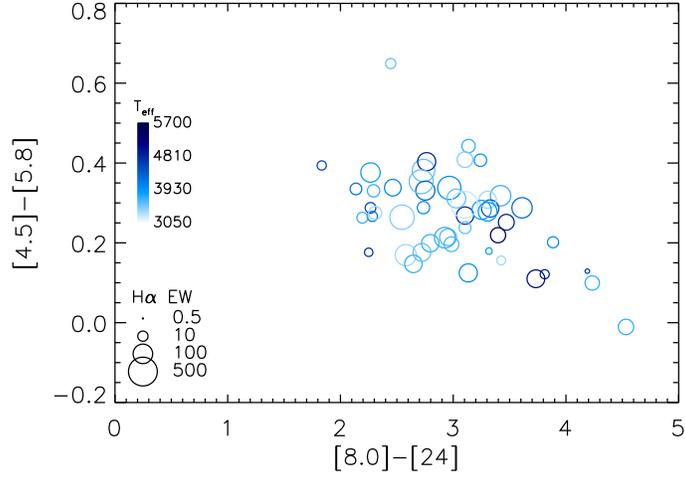}
\caption{IRAC+MIPS color-color diagram for those stars in our spectroscopic sample. Size of the symbol indicates strength of \ha, while color indicates $T_{eff}$, as in figure~\ref{irac_ccd_haew}.\label{irac_mips_haew}}
\end{figure}

An IRAC color-color diagram, Figure~\ref{irac_ccd_haew}, shows a similar progression in disk properties. There is a population of sources with no infrared excess, a population with strong infrared excess, and population in between with weak but significant infrared excess. Including the 24\micron\ band of MIPS adds additional information on the dust properties of the disk. Figure~\ref{irac_mips_haew} shows the [8.0]-[24] vs. [4.5]-[5.8] color-color diagram. All of the detections at 24\micron , 33/34 strong disks and  8/12 IRAC-weak disks, have some sort of disk emission as expected given the 24\micron\ sensitivity. Of the stars with an infrared excess not detected at 24\micron\ one is outside the MIPS field of view and the other three are in regions of strong nebulous emission where the sensitivity is decreased. 

\begin{figure}
\epsscale{.6}
\plotone{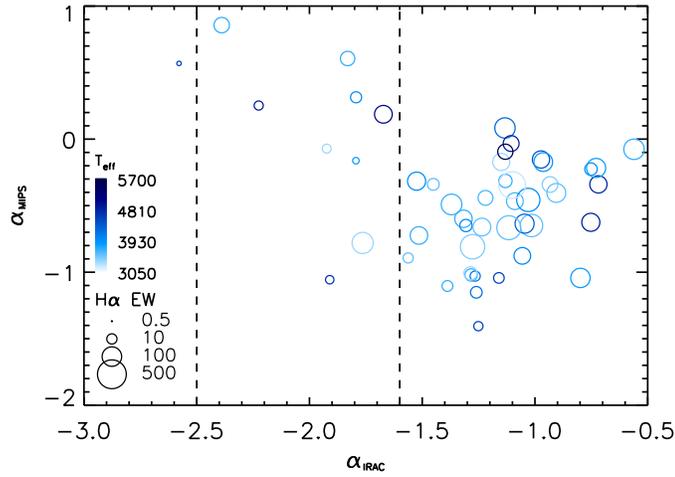}
\caption{SED slope for the IRAC bands versus the SED slope from 8-24\micron. The size of the points has been scaled to the size of the \ha\ equivalent width, while color scales with $T_{eff}$ as in figure~\ref{irac_ccd_haew}. The dashed lines delineates between strong disks, IRAC weak disks and stars with no disk. The MIPS weak sources are also evident around $\alpha_{MIPS}\approx-1$, and the transition disks have large $\alpha_{MIPS}$ and small \alphairac. The SEDs have not been dereddened.\label{alpha_IRAC_MIPS}}
\end{figure}

\begin{figure}
\epsscale{.6}
\plotone{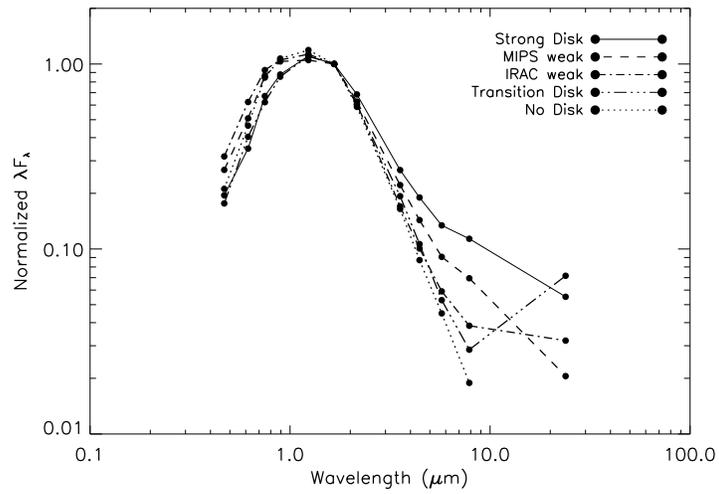}
\caption{SEDs of stars exhibiting different levels of infrared excess emission. SEDs have been dereddened using \citet{mat90} for $\lambda<3\micron$ and \citet{fla07} for $\lambda>3\micron$ and normalized to the H band flux. All of these stars have spectral type M1-M2.\label{drseds}}
\end{figure}

There is a population of sources with [8.0]-[24]$<$2.4, which we refer to as MIPS-weak. These sources are singled out because this [8.0]-[24] color indicates spectral slopes $<-1$, similar to $\lambda F_{\lambda}\propto\lambda^{-4/3}$, and because of the evidence for lower accretion activity, discussed below. A perfectly flat disk has an SED with the form $\lambda F_{\lambda}\propto\lambda^{-4/3}$, which represents the limit of a disk with a substantial amount of grain settling. Figure~\ref{alpha_IRAC_MIPS} shows \alphairac\ vs. $\alpha_{MIPS}$ , the SED slopes from $3-8\micron$ and $8-24\micron$ respectively. The MIPS-weak sources all have a strong excess at the IRAC bands, even though they have a weak excess at 24\micron. The lack of MIPS weak sources with small \alphairac\ may be caused by the limited sensitivity at 24\micron. The majority of sources show strong excesses at both the IRAC bands and at 24\micron, while the IRAC-weak sources have a range of excesses at 24\micron. Figure ~\ref{drseds} shows sample SEDs for each of the types of circumstellar disks. The biggest change in the SED between the MIPS weak star and the strong disk star is between 8 and 24\micron\ with the MIPS-weak disk having a much steeper SED beyond 6\micron. The SED of the IRAC-weak sources looks similar to the SED of the strong disk but with a steeper 2-8\micron\ slope.

Stars contaminated by background PAH emission may appear to have an excess at [8.0], even though they have no circumstellar material within a few AU. Three stars, \#802, 984, 1078, are in regions of high nebular emission and have an excess primarily at [8.0]. Excess emission at 24\micron\ would help to better characterize the presence of a disk, but none of these stars are detected at [24] possibly due to the high background emission surrounding the stars. We leave them as IRAC-weak disks but note that their disk status is still uncertain

Including all stars with \alphairac $>-2.5$ as having a disk as well as the stars with \alphairac$<-2.5$ but a 24\micron\ excess indicative of an optically thick outer disk, 79\% (53/67) of the spectroscopically confirmed members have disks. This disk fraction may not accurately reflect the disk fraction in the entire cluster, since sources with interesting mid-infrared colors were added to our spectroscopic sample. To constrain the true disk fraction we consider the photometry of those confirmed members, as well as stars with similar colors. In the r vs. r-i color-magnitude diagram (shown in Figure~\ref{sdss_cmd}), there are 397 stars that sit above the main sequence locus and have r$<18.6$, 230 of which have been observed spectroscopically. Of these 230, 44 are members with disks (\alphairac$>-2.5$) and 14 are members without disks. Our sample of members without disks is unbiased since we have no prior information on whether or not they are members. Scaling up from the spectroscopic sample to the photometric sample, we expect 24 stars without disks in the entire sample of 397 stars. If we assume that we have found all the members with disks, then out of the entire sample there would be 24 stars without a disk, and 44 stars with a disk for a disk fraction of 44/68 (65\%). We have likely not found all the members with disks, and the true disk fraction is likely higher than 65\%. Out of the 397 stars, 58 have infrared colors consistent with disk emission. If we instead assume that we have found all of the members without disks and that all of the objects with an infrared excess are members, then the disk fraction would be 58/72 (81\%). Some of the objects with an infrared excess may be background AGN or AGB stars and it is also unlikely that we have found all of the members without disks so we expect the true disk fraction to be less than 81\%. This does not count stars that could be missing material within 1 AU but the small number of these objects (see \S~\ref{disk_hole}) will not have a significant effect. The spectroscopic sample has a disk fraction of 79\%, while we expect the true fraction for the entire cluster to be in the range 65-81\%. 

Another estimate of the disk fraction can be made from the X-ray emitting stars observed by \citet{ski07}. X-ray emission as a tracer of young stars is not biased towards stars with disks. We detect 23 of the 33 X-ray emitting stars with IRAC, mainly at [3.6], [4.5]. We use these two colors, with a boundary of [3.6]-[4.5]=0.2 to differentiate stars with or without disks. One star is not detected by 2MASS, which combined with its faint IRAC flux suggests it is an extragalactic source and we exclude it. Where possible we use detections in other bands to better characterize the presence of a disk around these stars. Two stars with [3.6]-[4.5]$>0.2$ have colors consistent with highly reddened photospheres while two stars with [3.6]-[4.5]$<0.2$, which are both part of our spectroscpoic sample (\#984,1078), have a slight excess at [8.0]. In total there are 14/22 ($64\pm17\%$) stars with an infrared excess in this field. This is consistent with our previous estimate of the true disk fraction. Ten of these stars, six of which have an infrared excess, are deeply embedded and are not detected in the optical. The X-ray fields are in areas of high background emission at 24\micron\ reducing the sensitivity at this band and making it difficult to estimate the fraction of evolved disks in this field.

If the deeply embedded population has a higher disk fraction than our spectroscopic sample then this will introduce an additional bias in our measure of the cluster disk fraction. Ices are expected to form on dust grains within the molecular cloud above $A_v$=3 \citep{whit01}. We might expect that the stars with $A_v>$3 differ from the less embedded population because of the change in the surrounding molecular cloud. Of the stars with $A_v>3$, 93\% (28/30) have disks while 71\% (20/28) of the stars with $A_v<3$ have disks. This only includes stars earlier than M4; embedded stars later than M4 are too faint in the optical to be observed here. There does not appear to be a significant age difference between these two groups although the change in the disk fraction is suggestive of a difference between the embedded population and the revealed population. Observations of more heavily extincted cluster members are needed to characterize this difference fully.

\subsubsection{Disk Hole Sources\label{disk_hole}}
A number of stars have been observed towards other young clusters that appear to be missing dust within a few AU of the central star \citep{qui04,cal02,rice03}. These objects, often referred to as transition disks, lack infrared emission shortward of 10\micron\ but exhibit strong emission longwards of 10\micron. NGC 2068/71 has two such objects, objects 177 and 281. They have [8.0]-[24]$>4$ and [3.6]-[4.5]$<$0.1. Our high-resolution spectra allows a search for any weak accretion. Object 177 (FW 10\%=124 km/sec, \ha\ EW=1.2\AA) does not appear to be actively accreting, while object 281 has a broad, asymmetric \ha\ profile indicating active accretion (FW 10\%=400 km/sec, \ha\ EW=18.1\AA). A veiling of r=0.36 was measured for object 281 corresponding to an accretion rate of $\dot{M}=2.4\times10^{-9}$\accunits. This accretion rate is the smallest rate measured among the stars in this cluster, although our measured accretion rates are biased toward those stars whose accretion rate is high enough to have a measurable veiling.

\subsubsection{Infrared Excess and Accretion}
\begin{figure}
\epsscale{.6}
\plotone{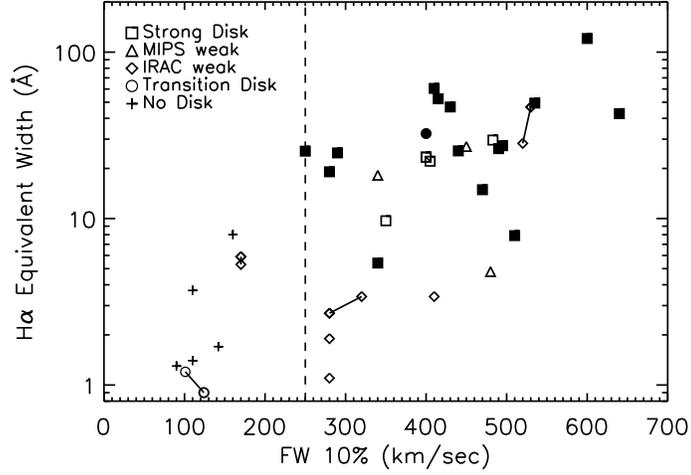}
\caption{Equivalent width of the \ha\ line versus the full width at 10\% maximum height for the \ha\ line for those likely members with high-resolution spectra. Vertical dashed line is located at 250 km/sec, our boundary for actively accreting stars. Different symbols represent different types of disks while filled symbols are stars with a measurable amount of veiling. Stars with multiple spectra have the two observations connected by a line\label{hafw}}
\end{figure}

 The stars with strong disks (32/67, 48\%), with spectral types ranging from G6-M5, all have signs of active accretion. They have strong \ha\ EW (5-300\AA) and broad asymmetric \ha\ profiles. The full width at 10\% maximum flux for the stars with strong disks ranges from 260 to 640 km/sec. For those stars with measurable veiling the accretion rates are in the range $7.9\times10^{-9}-8.6\times10^{-8}$\accunits. The stars with no disks (14/67, 21\%) show no sign of active accretion, either in \ha\ EW (1.3-9.5\AA) or the \ha\ profile (FW at 10\%=90-200 km/sec). Figure~\ref{hafw} shows the \ha\ EW versus FW at 10\% where the strong disks and stars with no disks form two distinct groups. 

The MIPS-weak sources (8/67, 12\%), with spectral types ranging from K4-M4, show signs of lowered accretion activity in addition to lower infrared excess. In the color-color diagrams, Figures~\ref{irac_mips_haew},~\ref{alpha_IRAC_MIPS}, the size of the symbols scales with the \ha\ EW and the MIPS-weak sources have smaller \ha\ EW than the strong disks. These eight stars have \ha\ EW between 3 and 48\AA. Three have high-resolution spectra with 10\% full-widths of 480, 340 and 450 km/sec, and accretion upper limits of $2.4\times10^{-8}$\accunits, $1.79\times10^{-8}$\accunits and $1.8\times10^{-7}$\accunits. In Figure~\ref{hafw} some of the MIPS-weak sources with high-resolution spectra are distinct from the strong disks and the stars with no disk. The \ha EW is lower than the strong disks with comparable FW at 10\%.

The IRAC-weak sources (12/67, 18\%, not including \#281), with spectral types ranging from K1-M4, have $-2.5<$\alphairac$<-1.6$ and also tend to have [8.0]-[24]$>2$. \citet{la06} find a similar spread in [8.0]-[24] color for their optically thin/''anemic'' disks. Of these twelve stars, six have high-resolution spectra that show 10\% full widths for \ha\ of 280, 410, 280, 260, 170 and 520 km/sec, close to the boundary between accretors and non-accretors. For these six objects only upper limits of $7.11,0.94,1.78,5.4,0.47,1.06\times10^{-8}$\accunits\ can be placed on the accretion rate since no veiling is measured. Figure~\ref{hafw} indicates they have the same accretion activity as the three MIPS-weak stars, separate from the strong disks and the stars with no disk. These stars have \ha\ EW between 2 and 32\AA, with one star having an \ha\ EW of 75\AA. With the exception of the one active accretor, these stars show less accretion activity than the strong disks.

\section{Discussion\label{discussion}}

Many studies now show that the disk fraction decreases rapidly with age, with nearly all stars losing their primordial disks after roughly 10 Myr \citep{hll01,her07}. For NGC 2068/71 we measure an age of $2\pm1.5$ Myr and a disk fraction of $79\%$. Our disk fraction is consistent with ground-based estimates from JHKL for regions such as the Trapezium and NGC 2024 with ages of 1-2 Myr \citep{hll01}.

In NGC 2068/71 the members with disks do not show a systematically different rotational velocity than those stars without disks. Rotational velocities, vsini, relative to the break-up speed are shown in Figure~\ref{vsini}, with the stars with and without disks marked as red and blue respectively. There appears to be no significant separation, although the substantial number of upper limits makes it difficult to evaluate any difference fully. It has been suggested that the slow rotation rates for pre-main sequence stars are due to disk-locking, where the star is connected to the disk by its magnetic field lines and angular momentum is transferred from the star to the disk. \citet{rws04} suggest that even if the slow rotation speed was due to disk locking, a significant difference between the rotation velocity of stars with disks and stars without disks would be difficult to measure. More data are needed to test these hypotheses in detail with the members of this region. 

The high-resolution spectra not only allow us to measure rotational velocities, but also obtain a more sensitive measure of the accretion activity with the \ha\ profile. All of the WTTS, classified based on \ha\ EW, that have a disk and high resolution spectra exhibit \ha\ profiles consistent with active accretion, except the transition disk \#177. Objects \#849, 458, 581 and 998 all have low-resolution \ha\ EW which would make them appear as WTTS, while their high-resolution \ha\ profiles indicate that they are still accreting material. This is especially important when examining the accretion activity of the evolved disks which tend to have low \ha\ EW although they are still accreting (see Figure~\ref{hafw}).

\begin{figure}
\epsscale{.6}
\includegraphics[angle=90,width=8in]{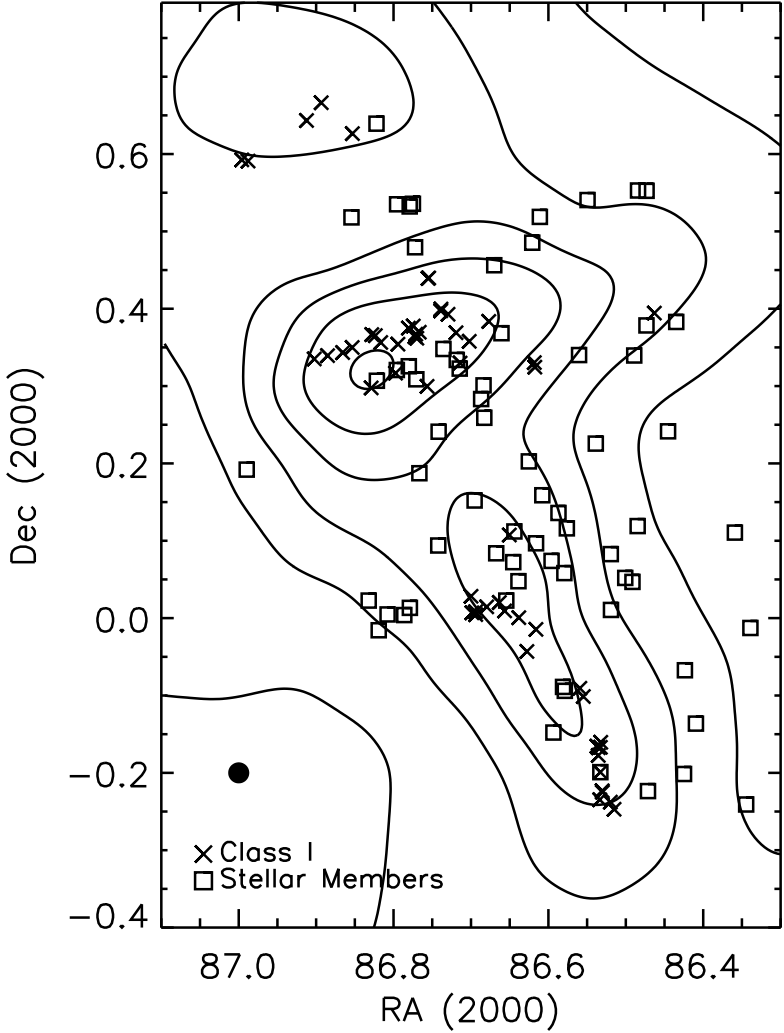}
\caption{Extinction map (axes in degrees) with class I sources from Muzerolle et al. (in prep) and the stellar members identified in this paper. The circle in the lower left shows the size of the gaussian kernel used to smooth the extinction map. The class I sources are located near the peaks in the extinction while the members follow the broader distribution of dust. Contours correspond to $A_V=1,2,3,4,5$ using the extinction law of \citet{rl85}\label{extmap}}
\end{figure}

The location of our selected members does not strictly overlap with either the dense gas emission \citep{lad91} or the dense dust emission \citep{joh01}. Star formation is ongoing in these regions of dense gas and dust, based on observable outflows and class I sources. To compare the distribution of members to the dust cloud we created an extinction map (Figure~\ref{extmap}) by selecting background stars, assuming they have the same intrinsic H-K$_s$ color, and calculating $E(H-K_s)=H-K_s-(H-K_s)_0$. A 30'' gaussian kernel was used to smooth the map and the extinction law of \citet{rl85} was used to convert from E(H-K$_s$) to $A_V$. A more detailed description of this method is given in \citet{goe03}. The extinction map closely matches the CS emission measured by \citet{lad91}. It also agrees with the lower resolution extinction map of \citet{dob05}, although their map does show a more extended peak at $\alpha_{2000}=05^h47^m36^s$, $\delta_{2000}=+00^{\circ}18'00''$ which could be due to our small number of background stars in this area. As seen in Figure~\ref{extmap}, our selected members follow the general shape of the diffuse dust traced by the extinction map, but are not confined to the peaks of the dust, where the SCUBA clumps \citep{joh01} and class I sources (Muzerolle et al. in prep) are located. Our sample may be biased toward those areas of lower extinction away from the dense gas and dust because the use of optical spectra for spectral type classification limits our survey to those young stars that are not deeply embedded within the cloud. Another possibility is that the star formation in this region has progressed from west to east. Stellar winds and photoionizing flux from the OB associations to the west of NGC 2068/71 may have been interacting with the cloud, causing stars to form on the western side of the cloud first and on the eastern side of the cloud later. A third possibility is that the selected members may have had time to disperse from their original birthplace and have spread throughout the cloud. For a typical velocity of 1-2 km/sec relative to the cloud, these members could have traveled 2-4 pc in their lifetime of ~2 Myr, roughly the east-west extent of the members.

Binarity may have a significant impact on disk properties, however, our data lack the spatial resolution to address this. One star (object \#739) was observed to be a binary by \citet{pad97}. This binary has a separation of 0.97'', which corresponds to 390 AU at a distance of 400pc. Interestingly this star has a MIPS-weak disk, which might be explained by tidal truncation of the outer disk. \citet{pad97} estimate a binary fraction of 15\%, which would imply that 10 of our sources are binaries. None of our stars with high-resolution spectra appear to be spectroscopic binaries but more epochs are needed to rule out single-line spectroscopic binaries.

\subsection{Evolved Disks}\label{evolved_disks}

There are a number of possible explanations for the SEDs of strong disks vs. MIPS-weak/IRAC-weak disks. Grain growth and settling is one process that may have a measurable effect on the SED of an accretion disk. Models suggest that this can be a very fast process producing noticeable differences in the SED in less than a million years \citep{dd05}. The overall effect is to quickly deplete grains out of the upper layers of the disk, while grains in the inner disk will grow and settle more quickly than in the outer disk.

As the degree of settling increases, the most direct effect is a decrease in the irradiation surface of the disk, which leads to less emerging flux. The SED approaches that of a perfectly flat disk, $\lambda F_{\lambda}\propto\lambda^{-4/3}$ at wavelengths corresponding to the disk radii that are affected by settling \citep{dal06}. Assuming the amount of grain growth and settling is constant throughout the disk, the change in flux is most noticeable for $\lambda>10\micron$ because these wavelengths trace regions in the disk where flaring becomes prominent and the surface height is most sensitive to grain growth and settling. The emission at $\lambda<10\micron$ is dominated by the wall at the dust destruction radius, which is less affected by the degree of flaring of the disk. For a substantial amount of grain growth and settling the height of the wall will decrease by a factor of 2, while the height of the outer disk can decrease by more than an order of magnitude \citep{dal06}. Theory predicts grain growth and settling should occur more rapidly in the inner disk than the outer disk since the orbital period is shorter \citep{dd04,dd05}. The IRAC-weak stars may have a substantial amount of grain growth and settling in the inner disk, whereas the MIPS-weak sources may have advanced settling over a wider range of radii.

The predicted SEDs for grain growth and settling are not unlike those corresponding to a decrease in the accretion rate. A lowered accretion rate will affect the disk in three ways: (1) decrease the surface density, (2) decrease the viscous heating and (3) decrease the accretion luminosity emitted by the stellar accretion shock. A decrease in surface density will lower the flux at all wavelengths, while the smaller viscous heating will lead to a cooler midplane temperature and a change in the scale height of the disk similar to that of grain growth and settling. The decrease in the accretion luminosity will have an effect on the inner disk edge which dominates the flux at $\lambda<10\micron$. Accretion luminosity, along with stellar irradiation, illuminates the inner dust edge of the disk. A decrease in accretion luminosity will cause the dust destruction radius to move closer to the star. While the inner edge will still be at 1500K, its solid angle will decrease and the emission from the inner edge will decrease. For $\dot{M}>10^{-8}$\accunits\ the accretion luminosity is comparable to the stellar luminosity and a change in accretion rate leads to a change in the radius of the wall, and a change in the flux in the IRAC bands. Since the accretion luminosity ($L=GM\dot{M}/R$) and the stellar luminosity roughly scale with stellar mass, the dependence on accretion rate dominates the ratio of accretion to stellar luminosity. For $\dot{M}<10^{-8}$\accunits\ the change in the radius of the wall with accretion rate is smaller because the stellar luminosity dominates.

A change in the SED slope could be due to grain growth and settling or a decreased accretion rate separately or it could be caused by a combination of the two. As the amount of settling increases the effect of a change in the accretion rate becomes more significant \citep{dal06}. For high accretion rates, viscous heating dominates over other forms of heating in the surface layers. Since viscous heating has little dependence on the grains in these layers, there is little change in the SED when grains start to grow. For lower accretion rates, stellar irradiation dominates the heating of the surface layers, and this form of heating is very sensitive to the grain distribution. The MIPS-weak and IRAC-weak sources may be separating themselves from the strong disks because their low accretion rates lead to a more significant difference in the SED. However, we need more estimates of $\dot{M}$ to directly compare to disk models.

This picture is confused by the $10\micron$ silicate feature which overlaps with the $8\micron$ band of IRAC. A strong silicate emission feature will artificially increase \alphairac\ despite a steeper continuum slope. As the accretion rate decreases the intensity of this feature increases due to the contrast in temperature between the surface (heated by stellar irradiation) and the midplane (heated by viscous dissipation and emission from the upper layers) \citep{dal06}. The silicate feature is most sensitive to grain growth in the upper layers of the disk and does not trace grain growth in the midplane. Spectroscopy of the $10\micron$ silicate feature would help to break the degeneracy between grain growth and settling and a change in accretion rate \citep{dal06,fur06}, as well as seperate out its contibution to \alphairac.

Another possibility for the observed SEDs of the MIPS-weak stars is the truncation of the disk by a close binary. A close binary will truncate the disk around the primary at 1/2 to 1/3 the semimajor axis of the binary system \citep{art94}. The majority of the 24\micron\ emission comes from within 0.5-5 AU, depending on the spectral type and degree of settling in the disk \citep{dal06}. A binary companion separated by $\sim$1-15AU would truncate the disk and reduce the 24\micron\ emission without affecting the shorter wavelength excess. The frequency of MIPS-weak stars in NGC 2068/71 is roughly consistent with the frequency of binaries at separations $\sim$1-15AU \citep{mat94}. The smaller resevoir of material available to accrete onto the central star may lead the accretion rate to decrease earlier than it would for stars with disks that have not been truncated. The smaller disks imply a smaller viscous timescale for the evolution of the disk, which may lead it to be quickly dissipated unless it is ressuplied by a circumbinary disk. The binary fraction within a cluster seems to be a function of environment, rather than of age, with low-desnity clusters such as Taurus and Ophiuchus showing a higher binary fraction than dense clusters such as the Trapezium \citep{mat00}. Millimeter observations find that disk masses are much lower for binaries separated by 1-100 AU than single stars or wide binaries, suggesting truncation of the disk \citep{mat00}. If MIPS-weak disks are due to the presence of a binary then we may expect the MIPS-weak fraction to be a strong function of environment rather than age. Interestingly one of our MIPS-weak stars (\#739) is a known binary \citep{pad97}, although its separation (390AU) is likely to be too large for any truncation of the disk to affect the 24\micron\ flux.

The third group of evolved disks are the transition disks, with AU-scale optically thin or evacuated holes in the center of the disks. We have two such objects, \#177, and 281, making up 3\% of our spectroscopic sample. One appears to be actively accreting (\#281) while the other is not. \citet{sa06}, who define transition disks as having no excess shortward of 6\micron, find that half of the transition disks in Tr 37 are still actively accreting, based on high-resolution \ha\ profiles. The transition disks around TW Hya, GM Aur and DM Tau are all seen to be accreting \citep{cal02,cal05} while CoKu Tau/4 is not actively accreting \citep{dal05}. Accretion rates of $10^{-8}-10^{-10}$\accunits\ have been measured for GM Aur, DM Tau, and TW Hya \citep{gull98,cal05,mul00}, similar to that measured for object \#281.

A hole could be swept out by a planet, be caused by substantial grain growth in the inner disk or be due to photoevaporation of the disk. The formation of a planet would cause material outside the planet's orbit to gain orbital angular momentum and be pushed outwards, while material within the planets orbit would lose angular momentum and quickly accrete onto the star \citep{rice03}. The mass of the planet and the disk will affect whether a gap can be created, and how the planet reacts to the material in the outer disk \citep{pap07}. A planet has been invoked to explain the observed inner holes in GM Aur \citep{rice03}, CoKu Tau/4 \citep{qui04} and TW Hya \citep{cal02}. For object \# 281, the material within the planet's radius may not have completely accreted onto the star, or the planet may be small enough that some gas may be able to stream past it \citep{lub06}. This gas would transport small grains into the gap, creating an optically thin disk, which would not be detectable from photometry alone \citep{rice06}. It is also possible that grains within 10 AU have grown to millimeter sizes or larger, while the grains at larger radii have not grown as much. The growth of grains should proceed on an orbital timescale, which would result in more grain growth in the inner disk than the outer disk \citep{dd04,dd05}, although this process should not produce a sharp boundary in the dust disk. Any accretion onto the central star may also transport small grains back into this gap. A third possibility arises from a photoevaporative wind from the disk driven by stellar UV and X-ray photons \citep{cla01,ale06}. When the accretion rate is low the wind will counteract the inward flow of material outside of the gravitational radius and will prevent additional gas from being transported into the inner disk. With the inner disk starved for material any gas will quickly accrete onto the star and disappear producing an inner hole for a short time before the entire disk dissipates. This process requires very low ($<10^{-10}$\accunits) accretion rates in order for the wind to counteract the inward flow of material due to accretion. This would rule out the photoevaporative model for object \#281, but it is still possible for object \#177. 

None of these three groups of evolved disks appear to be systematically older than the rest of the cluster, according to the HR diagram shown in Figure~\ref{hr}. Although the two transition disks are below the 3 Myr isochrone their small number make any statistically significant conclusion impossible. While the number of sources in each group is small enough that we cannot do a detailed study of the ages, we can say that it is unlikely that either the IRAC-weak, MIPS-weak or transition disks are systematically older or younger than the other stars in the cluster. This indicates that the disks around different stars evolve at different rates. \citet{dd05} find that grain growth proceeds much more quickly than the typical age of a T-Tauri star and small grains must be replenished in the disk in order to retain a significant IR excess. If replenishment occurs then the typical grain size in a disk does not reflect the age of the system. We would then not expect these stars to appear systematically older if the change in SED for the IRAC-weak and MIPS-weak disks is due in part to grain growth. 

IRAC-weak disks are observed in cluster regions from Taurus at 1 Myr \citep{la06} to NGC 2362 at 5 Myr \citep{dah07}. The fraction of IRAC-weak disks depends on the spectral type range under consideration. \citet{la06} find the fraction of stars with IRAC-weak disks ranges from 8-25\% depending on the spectral type. Later spectral types have a larger fraction of IRAC-weak disks although this trend may be affected by larger photometric uncertainties at later spectral types blurring the boundary between IRAC-weak and diskless stars \citep{her07}. We examine the fraction of disked stars that have an IRAC-weak SED for broad spectral type ranges to look for trends with age. Here we use \alphairac=-1.8 as the boundary for IRAC-weak stars to be consistent with previous studies. We consider K0-M1 stars in Taurus at 1 Myr \citep{har05}, NGC 2068/71 at 2 Myr, IC 348 at 2-3 Myr \citep{la06}, $\sigma$ Ori at 3 Myr \citep{her07}, Tr 37 at 4 Myr \citep{sa06} and NGC 2362 at 5 Myr \citep{dah07}. The fractions of disked stars with IRAC-weak SEDs are listed in Table ~\ref{evolved_fraction}. The fractions of cluster members with disks in these clusters decrease from 80\%\ to $\sim20\%$ with age, but the fractions of disks with IRAC-weak SEDs do not show an obvious decrease with age, although the small number statistics for some of these clusters make this trend uncertain. Typical \ha\ EW for the IRAC-weak disks in Tr 37 and IC 348 are $<15\AA$ \citep{sa06,la06}, consistent with the IRAC-weak stars in NGC 2068/71, and available high-resolution \ha\ profiles indicate they are actively accreting. However \citet{her07} find that the median SED slope for all disks in a given region decreases with age from 1-3 Myr, which may contribute to the fraction of disks with IRAC-weak SEDs if the entire distribution of infrared excess is shifting to a smaller excess.

Of the 54 stars with disks detected by MIPS in Tr 37, 7 have MIPS-weak disks \citep{sa06}, while out of the 38 K0-M4 stars with disks in IC 348, 3 have MIPS-weak disks, none of which are earlier than M1 \citep{la06}. In $\sigma$ Ori, three out of 16 K0-M1 stars with disks detected by MIPS have MIPS-weak disks \citep{her07}. The sensitivity limit at 24\micron\ makes any trends with age harder to evaluate although we note that the fraction of disks with MIPS-weak SEDS does not significantly decrease with age. The \ha\ EW of the MIPS-weak stars in Tr 37 are between 4 and 47\AA, with most below 15\AA\ \citep{sa06}. Of the stars with measured accretion rates, most are below $10^{-8}$\accunits\ \citep{sa06}. This is consistent with the properties of MIPS-weak disks in NGC 2068/71, although overall accretion rates are lower in Tr 37 and the low accretion rates for the MIPS-weak stars may simply reflect viscous evolution of the disks.

If disk evolution is not solely dependent on age then it may depend on environment or the initial conditions of the protostar. Material infalling onto the disk from the surrounding cloud could replenish the small grains in the disk as well as maintain the presence of a disk. The longevity of a disk would then depend on the resevoir of material surrounding the star. During the early stages of star formation material accretes from an envelope onto the disk, and this infall rate depends on the properties of the envelope. More material accreting onto the disk will create a more massive disk, which can affect its subsequent evolution \citep{dd05}. Unfortunately sub-mm observations are not sensitive to circumstellar disks in NGC 2068/71and cannot be used to estimate disk masses for these stars \citep{joh01}. The initial angular momentum of the system may influence the initial size of the disk and how quickly it evolves \citep{dul06} as well as affect the possibility of forming a binary, which could enhance the subsequent evolution of the disk. A nearby O/B star would increase the ionizing flux impinging on the surface of the disk, and possibly increase the speed with which the disk dissipates, although the evolved disks we observe do not appear preferentially close to the B stars in this cluster and with the earliest member being a B2 star the UV flux is not especially strong.

\section{Conclusion}
We have studied the young stellar population of NGC 2068/71 and identified 69 members, for 67 of which we have derived spectral type and luminosity. Our cluster members range from G6-M6 with a median age of 2 Myr. A large fraction (79\%) of these stars have an infrared excess, although this is likely an overestimate of the true disk fraction. Of the stars with an infrared excess and a high-resolution \ha\ profile, sensitive to low-accretion rates, all but one are actively accreting. The stellar members are not confined to regions of dense gas and dust, although there is some evidence that the disk fraction is higher in dense ($A_V>3$) regions.

Combining accretion information from the \ha\ line with the 3-24\micron\ infrared SED we find three populations of evolved disks: IRAC-weak, MIPS-weak and transition disks. All of the populations show a change in the shape of the infrared SED as well as a decrease in the accretion activity. For the IRAC-weak disks this could solely be due to a decrease in the accretion rate (which affects the structure of the disk), substantial grain growth and settling, or a combination of the two. These two processes could also explain the MIPS-weak SEDS, although truncation of the outer disk by a close binary companion is also a viable possibility. The transition disks could be explained by a giant planet opening a gap in the disk, or a photoevaporative wind evacuating the inner disk. The latter mechanism requires a very low accretion rate and can be ruled out for the transition disk in our sample that is still accreting.

None of the groups of evolved disks appear systematically older than the rest of the cluster, although the small numbers make any difference harder to evaluate. Initial conditions in the disk and surrounding envelope could influence how quickly disks evolve. The initial angular momentum or mass of the disk could also have an influence.

\acknowledgments
This publication makes use of data products from the Two Micron All Sky Survey, which is a joint project of the University of Massachusetts and the Infrared Processing and Analysis Center/California Institute of Technology, funded by the National Aeronautics and Space Administration and the National Science Foundation. This work is based in part on observations made with IRAC, aboard the {\it Spitzer Space Telescope}, and is supported by contract \#1255094 for SSC/Caltech to the University if Arizona. The Second Palomar Observatory Sky Survey (POSS-II) was made by the California Institute of Technology with funds from the National Science Foundation, the National Geographic Society, the Sloan Foundation, the Samuel Oschin Foundation, and the Eastman Kodak Corporation. We would also like to thank Tom Megeath for providing the IRAC data and George Rieke for comments that helped improve the paper.

\begin{figure}
\epsscale{.6}
\caption{POSS blue plate centered at $\alpha_{2000}=05^h46^m47^s$ $\delta_{2000}=+00^{\circ}09'50''$ spanning one degree on each side overlayed with the positions of cluster members.\label{poss_image}}
\end{figure}

\begin{deluxetable}{ccccccccccccc}
\rotate
\tabletypesize{\footnotesize}
\tablewidth{0pt}
\tablecaption{Photometry of likely members\label{member_phot}}
\tablehead{\colhead{Object ID} & \colhead{g} & \colhead{r} & \colhead{i} & \colhead{z} & \colhead{J} & \colhead{H} & \colhead{$K_s$} & \colhead{[3.6]} & \colhead{[4.5]} & \colhead{[5.8]} & \colhead{[8.0]} & \colhead{[24]} } 
\startdata
  17 & 18.723 & 17.104 & 15.488 & 14.567 & 12.962 & 12.217 & 11.932 & 11.697 $\pm$ 0.009 & 11.612 $\pm$ 0.011 & 11.611 $\pm$ 0.027 & 11.515 $\pm$ 0.033 & \nodata \\
 29 & 19.946 & 18.352 & 16.403 & 15.281 & 13.497 & 12.874 & 12.541 & 12.279 $\pm$ 0.011 & 12.153 $\pm$ 0.014 & 12.167 $\pm$ 0.040 & 12.077 $\pm$ 0.047 & \nodata \\
 63 & 17.635 & 16.174 & 14.778 & 13.952 & 12.371 & 11.606 & 11.208 & 10.619 $\pm$ 0.005 & 10.318 $\pm$ 0.005 & 10.006 $\pm$ 0.014 & 9.252 $\pm$ 0.007 & \nodata \\
 141 & 18.160 & 16.337 & 15.190 & 14.415 & 12.645 & 11.820 & 11.526 &11.356 $\pm$ 0.005 & 11.329 $\pm$ 0.008 & 11.319 $\pm$ 0.021 & 11.233 $\pm$ 0.022 & \nodata \\
 174 & 18.373 & 16.823 & 15.297 & 14.436 & 12.863 & 12.092 & 11.855 &11.584 $\pm$ 0.008 & 11.528 $\pm$ 0.009 & 11.597 $\pm$ 0.030 & 11.503 $\pm$ 0.030 & \nodata \\
 177 & 15.897 & 14.469 & 13.808 & 13.378 & 11.925 & 11.183 & 10.989 &10.862 $\pm$ 0.006 & 10.947$\pm$ 0.007 & 10.818 $\pm$ 0.024 & 10.667 $\pm$ 0.017 & 6.478 $\pm$ 0.036 \\
 204 & 17.221 & 15.528 & 14.564 & 13.862 & 12.021 & 11.051 & 10.376 & 9.387 $\pm$ 0.003 & 8.943 $\pm$ 0.003 & 8.744 $\pm$ 0.007 & 8.006 $\pm$ 0.004 & 5.209 $\pm$ 0.033 \\
 229 & 19.843 & 18.279 & 16.606 & 15.638 & 14.005 & 13.355 & 13.044 & 12.806 $\pm$ 0.014 & 12.716 $\pm$ 0.016 & 12.781 $\pm$ 0.056 & 12.740 $\pm$ 0.066 & \nodata \\
 281 & 17.718 & 16.147 & 15.253 & 14.508 & 12.777 & 11.956 & 11.694 & 11.362 $\pm$ 0.006 & 11.280 $\pm$ 0.009 & 11.291 $\pm$ 0.021 & 10.942 $\pm$ 0.022 & 6.410 $\pm$ 0.032 \\
 285 & 18.420 & 16.458 & 15.304 & 14.423 & 12.452 & 11.404 & 10.870 & 10.201 $\pm$ 0.004 & 9.857 $\pm$ 0.006 & 9.504 $\pm$ 0.009 & 8.686 $\pm$ 0.007 & 5.970 $\pm$ 0.032 \\
 288 & 16.758 & 15.346 & 14.685 & 14.275 & 13.042 & 12.335 & 12.153 & 12.067 $\pm$ 0.008 & 12.066 $\pm$ 0.012 & 12.047 $\pm$ 0.029 & 11.956 $\pm$ 0.038 & \nodata \\
 322 & 19.394 & 17.911 & 15.799 & 14.694 & 12.824 & 12.252 & 11.921 & 11.601 $\pm$ 0.008 & 11.476 $\pm$ 0.010 & 11.450 $\pm$ 0.028 & 11.382 $\pm$ 0.029 & \nodata \\
 326 & 16.791 & 15.137 & 14.247 & 13.641 & 12.133 & 11.331 & 11.027 & 10.167 $\pm$ 0.004 & 9.921 $\pm$ 0.004 & 9.719 $\pm$ 0.012 & 9.241 $\pm$ 0.008 & 5.355 $\pm$ 0.031 \\
 344 & 16.564 & 15.029 & 14.187 & 13.517 & 11.754 & 10.825 & 10.215 & 9.291 $\pm$ 0.003 & 9.009 $\pm$ 0.004 & 8.885 $\pm$ 0.008 & 8.099 $\pm$ 0.005 & 4.965 $\pm$ 0.032 \\
 351 & 19.240 & 17.774 & 16.328 & 15.258 & 13.517 & 12.677 & 12.152 & 11.372 $\pm$ 0.007 & 11.134 $\pm$ 0.008 & 10.965 $\pm$ 0.022 & 10.412 $\pm$ 0.018 & 7.832 $\pm$ 0.034 \\
 373 & 18.355 & 16.627 & 14.854 & 13.850 & 12.047 & 11.218 & 10.905 & 10.424 $\pm$ 0.005 & 10.110 $\pm$ 0.004 & 9.802 $\pm$ 0.013 & 8.933 $\pm$ 0.006 & 5.627 $\pm$ 0.031 \\
 416 & 20.237 & 18.009 & 16.459 & 15.365 & 13.362 & 12.278 & 11.739 & 11.055 $\pm$ 0.005 & 10.752 $\pm$ 0.007 & 10.514 $\pm$ 0.013 & 9.821 $\pm$ 0.015 & 6.716 $\pm$ 0.032 \\
 418 & 17.062 & 15.362 & 14.480 & 13.642 & 11.339 & 10.125 & 9.455 & 8.296 $\pm$ 0.002 & 7.887 $\pm$ 0.003 & 7.550 $\pm$ 0.003 & 6.692 $\pm$ 0.003 & 3.726 $\pm$ 0.033 \\
 458 & \nodata & \nodata & \nodata & \nodata & 9.795 & 8.647 & 8.399 & 8.106 $\pm$ 0.003 & 7.974 $\pm$ 0.003 & 7.798 $\pm$ 0.004 & 7.292 $\pm$ 0.003 & 5.042 $\pm$ 0.033 \\
 480 & 16.846 & 15.135 & 14.963 & 13.377 & 11.686 & 10.698 & 10.054 &  8.947 $\pm$ 0.003 & 8.603 $\pm$ 0.003 & 8.161 $\pm$ 0.004 & 7.468 $\pm$ 0.004 & 4.332 $\pm$ 0.033 \\[10pt]
 515 & 14.447 & 13.147 & 12.620 & 12.295 & 10.936 & 10.046 & 9.448 &  8.680 $\pm$ 0.003 & 8.581 $\pm$ 0.003 & 8.460 $\pm$ 0.005 & 8.142 $\pm$ 0.005 & 4.330 $\pm$ 0.033 \\
 543 & 20.419 & 18.517 & 16.689 & 15.636 & 13.825 & 12.889 & 12.522 & 12.199 $\pm$ 0.010 & 11.965 $\pm$ 0.011 & 11.809 $\pm$ 0.029 & 11.380 $\pm$ 0.032 & 7.955 $\pm$ 0.032 \\
 571 & 18.808 & 16.447 & 15.165 & 14.279 & 12.208 & 11.068 & 10.471 & 9.534 $\pm$ 0.004 & 9.085 $\pm$ 0.003 & 8.682 $\pm$ 0.008 & 7.690 $\pm$ 0.009 & 4.925 $\pm$ 0.033 \\
 581 & 16.797 & 15.016 & 16.570 & 13.143 & 11.198 & 10.165 & 9.724 & 8.945 $\pm$ 0.004 & 8.654 $\pm$ 0.002 & 8.261 $\pm$ 0.006 & 7.569 $\pm$ 0.003 & 5.735 $\pm$ 0.031 \\
 584 & 18.588 & 16.103 & 14.552 & 13.762 & 10.632 & 9.030 & 7.869 &  6.542 $\pm$ 0.001 & 6.005 $\pm$ 0.001 & 5.674 $\pm$ 0.002 & 4.936 $\pm$ 0.002 & 2.184 $\pm$ 0.033 \\
 590 & 15.605 & 13.962 & 12.894 & 12.572 & 10.690 & 9.655 & 8.999 & 7.960 $\pm$ 0.002 & 7.497 $\pm$ 0.002 & 7.229 $\pm$ 0.004 & 6.051 $\pm$ 0.002 & 2.946 $\pm$ 0.031 \\
 601 & 19.965 & 17.488 & 15.935 & 14.912 & 12.533 & 11.322 & 10.685 & 9.923 $\pm$ 0.003 & 9.500 $\pm$ 0.004 & 9.094 $\pm$ 0.008 & 8.082 $\pm$ 0.006 & 4.841 $\pm$ 0.034 \\
 618 & 15.905 & 14.446 & 13.656 & 13.137 & 11.510 & 10.654 & 10.214 & 9.542 $\pm$ 0.003 & 9.365 $\pm$ 0.003 & 9.255 $\pm$ 0.008 & 8.491 $\pm$ 0.005 & 4.758 $\pm$ 0.031 \\
 628 & 20.090 & 17.977 & 16.367 & 15.222 & 13.061 & 11.974 & 11.429 & 10.786 $\pm$ 0.005 & 10.398 $\pm$ 0.005 & 10.017 $\pm$ 0.013 & 9.182 $\pm$ 0.019 & 6.446 $\pm$ 0.037 \\
 656 & 19.721 & 17.793 & 16.139 & 15.135 & 13.190 & 12.186 & 11.643 & 10.847 $\pm$ 0.005 & 10.373 $\pm$ 0.006 & 10.197 $\pm$ 0.013 & 9.382 $\pm$ 0.021 & 6.660 $\pm$ 0.036 \\
 663 & 19.260 & 16.916 & 15.516 & 14.501 & 12.695 & 11.330 & 10.517 & 9.546 $\pm$ 0.003 & 9.163 $\pm$ 0.004 & 8.897 $\pm$ 0.007 & 8.141 $\pm$ 0.005 & 5.858 $\pm$ 0.031 \\
 677 & 20.403 & 18.209 & 16.280 & 15.014 & 12.714 & 11.581 & 11.110 & 10.794 $\pm$ 0.004 & 10.725 $\pm$ 0.007 & 10.699 $\pm$ 0.021 & 10.613 $\pm$ 0.045 & \nodata \\
 689 & 18.100 & 16.168 & 14.854 & 13.969 & 12.170 & 11.298 & 10.992 & 10.755 $\pm$ 0.005 & 10.703 $\pm$ 0.006 & 10.671 $\pm$ 0.015 & 10.593 $\pm$ 0.017 & \nodata \\
 699 & 20.568 & 18.163 & 16.493 & 15.044 & 12.642 & 11.379 & 10.718 & 9.840 $\pm$ 0.004 & 9.521 $\pm$ 0.004 & 9.324 $\pm$ 0.010 & 8.384 $\pm$ 0.005 & 5.400 $\pm$ 0.032 \\
 712 & 17.662 & 15.846 & 14.773 & 13.721 & 11.632 & 10.441 & 9.627 & 8.445 $\pm$ 0.003 & 7.887 $\pm$ 0.002 & 7.466 $\pm$ 0.004 & 6.440 $\pm$ 0.002 & 3.367 $\pm$ 0.031\\
 730 & 18.805 & 16.620 & 15.322 & 14.266 & 12.715 & 11.172 & 10.213 & 8.838 $\pm$ 0.003 & 8.374 $\pm$ 0.002 & 8.036 $\pm$ 0.005 & 7.251 $\pm$ 0.005 & 4.787 $\pm$ 0.032 \\
 739 & 20.094 & 17.561 & 15.994 & 14.744 & 11.963 & 10.569 & 9.733 & 8.735 $\pm$ 0.003 & 8.266 $\pm$ 0.003 & 7.978 $\pm$ 0.006 & 7.230 $\pm$ 0.009 & 4.964 $\pm$ 0.034 \\
 741 & 20.192 & 18.130 & 16.165 & 14.957 & 12.706 & 11.563 & 11.101 & 10.752 $\pm$ 0.006 & 10.614 $\pm$ 0.007 & 10.325 $\pm$ 0.026 & 9.884 $\pm$0.052 & \nodata \\
 760 & 17.699 & 15.797 & 14.722 & 13.766 & 11.672 & 10.420 & 9.525 & 8.226 $\pm$ 0.002 & 7.791 $\pm$ 0.003 & 7.504 $\pm$ 0.004 & 6.701 $\pm$ 0.008 & 3.090 $\pm$ 0.032 \\
 782 & 20.679 & 18.266 & 16.655 & 15.474 & 13.411 & 12.192 & 11.703 & 11.137 $\pm$ 0.006 & 10.842 $\pm$ 0.007 & 10.579 $\pm$ 0.018 & 9.853 $\pm$ 0.011 & 7.659 $\pm$ 0.032 \\
 802 & 19.386 & 17.020 & 15.688 & 14.729 & 12.779 & 11.767 & 11.400 & 11.085 $\pm$ 0.007 & 11.038 $\pm$ 0.008 & 10.916 $\pm$ 0.036 & 10.520 $\pm$ 0.090 & \nodata \\[10pt]
 813 & 19.992 & 18.169 & 16.398 & 15.131 & 12.897 & 12.038 & 11.386 & 10.275 $\pm$ 0.003 & 9.866 $\pm$ 0.004 & 9.602 $\pm$ 0.008 & 8.877 $\pm$ 0.006 & 6.331 $\pm$ 0.031 \\
 843 & 19.311 & 17.150 & 16.020 & 14.773 & 12.932 & 11.926 & 11.385 & 10.351 $\pm$ 0.004 & 10.394 $\pm$ 0.005 & 9.745 $\pm$ 0.010 & 9.357 $\pm$ 0.008 & 6.911 $\pm$ 0.031 \\
 849 & 17.078 & 15.131 & 14.314 & 13.275 & 11.562 & 10.608 & 10.218 & 9.433 $\pm$ 0.004 & 9.130 $\pm$ 0.004 & 8.795 $\pm$ 0.006 & 8.049 $\pm$ 0.005 & 5.912 $\pm$ 0.032 \\
 856 & 20.608 & 18.157 & 16.730 & 15.309 & 13.190 & 12.057 & 11.570 & 11.001 $\pm$ 0.006 & 10.820 $\pm$ 0.007 & 10.721 $\pm$ 0.021 & 10.091 $\pm$ 0.024 & 5.858 $\pm$ 0.033 \\
 878 & 18.967 & 16.961 & 16.075 & 14.712 & 12.292 & 11.127 & 10.395 & 9.613 $\pm$ 0.004 & 9.073 $\pm$ 0.003 & 8.791 $\pm$ 0.013 & 7.708 $\pm$ 0.025 & 4.456 $\pm$ 0.036 \\
 934 & 19.906 & 17.685 & 16.486 & 15.120 & 13.290 & 12.302 & 11.838 & 11.232 $\pm$ 0.007 & 10.929 $\pm$ 0.007 & 10.715 $\pm$ 0.024 & 9.664 $\pm$ 0.022 & 6.710 $\pm$ 0.043 \\
 941 & 18.562 & 16.677 & 15.929 & 14.651 & 12.658 & 11.482 & 10.802 & 9.739 $\pm$ 0.003 & 9.314 $\pm$ 0.004 & 9.036 $\pm$ 0.008 & 8.058 $\pm$ 0.015 & 4.753 $\pm$ 0.034 \\
 984 & \nodata & \nodata & \nodata & \nodata & 11.937 & 10.917 & 10.557 & 10.341 $\pm$ 0.004 & 10.302 $\pm$ 0.005 & 10.220 $\pm$ 0.016 & 9.992 $\pm$ 0.050 & \nodata \\
 997 & 20.356 & 18.400 & 17.026 & 15.316 & 13.009 & 12.102 & 11.644 & 10.658 $\pm$ 0.005 & 10.264 $\pm$ 0.005 & 9.856 $\pm$ 0.013 & 8.985 $\pm$ 0.017 & 5.881 $\pm$ 0.033 \\
 998 & 18.092 & 15.752 & 14.349 & 13.290 & 11.230 & 10.128 & 9.690 & 9.395 $\pm$ 0.004 & 9.200 $\pm$ 0.003 & 9.021 $\pm$ 0.010 & 8.469 $\pm$ 0.021 & 5.152 $\pm$ 0.032 \\
1062 & 18.392 & 16.573 & 15.388 & 14.337 & 12.460 & 11.405 & 10.803 & 10.074 $\pm$ 0.004 & 9.649 $\pm$ 0.004 & 9.338 $\pm$ 0.010 & 8.349 $\pm$ 0.005 & 5.321 $\pm$ 0.031 \\
1078 & 19.855 & 17.001 & 15.631 & 14.260 & 11.839\tablenotemark{a} & 10.882 & 10.449 & 10.040 $\pm$ 0.005 & 9.890 $\pm$ 0.006 & 9.811 $\pm$ 0.029 & 9.496$ \pm$ 0.072 & \nodata \\
1089 & 20.339 & 18.208 & 16.573 & 15.556 & 13.720 & 12.733 & 12.409 & 12.084 $\pm$ 0.009 & 12.026 $\pm$ 0.012 & 12.040 $\pm$ 0.030 & 11.898 $\pm$ 0.038 & \nodata \\
1099 & 19.666 & 17.182 & 15.830 & 14.714 & 11.745 & 10.403 & 9.368 & 8.095 $\pm$ 0.003 & 7.578 $\pm$ 0.002 & 7.327 $\pm$ 0.004 & 6.526 $\pm$ 0.002 & 3.055 $\pm$ 0.032 \\
1116 & 15.603 & 14.124 & 13.451 & 12.932 & 11.350 & 10.570 & 10.223 & 9.817 $\pm$ 0.003 & 9.582 $\pm$ 0.004 & 9.296 $\pm$ 0.008 & 8.170 $\pm$ 0.005 & 4.842 $\pm$ 0.032 \\
1117 & 20.262 & 18.223 & 16.715 & 15.522 & 12.858 & 11.688 & 10.886 & 10.138 $\pm$ 0.004 & 9.502 $\pm$ 0.004 & 9.184 $\pm$ 0.008 & 8.078 $\pm$ 0.005 & 4.658 $\pm$ 0.031 \\
1119 & 17.576 & 15.067 & 13.842 & 12.556 & 10.226 & 9.018 & 8.324 & 7.399 $\pm$ 0.001 & 6.882 $\pm$ 0.002 & 6.506 $\pm$ 0.002 & 5.582 $\pm$ 0.004 & 3.316 $\pm$ 0.033 \\
1134 & 17.306 & 15.610 & 14.335 & 13.373 & 11.558 & 10.798 & 10.515 & 10.281 $\pm$ 0.004 & 10.196$ \pm$ 0.005 & 10.192 $\pm$ 0.015 & 10.070 $\pm$ 0.015 & \nodata \\
1171 & 14.938 & 13.031 & 12.562 & 12.373 & 12.563 & 11.642 & 11.191 & 10.404 $\pm$ 0.004 & 10.035 $\pm$ 0.005 & 9.705 $\pm$ 0.009 & 9.032 $\pm$ 0.008 & 6.739 $\pm$ 0.031 \\
1173 & 17.355 & 15.194 & 14.182 & 13.028 & 10.998 & 9.836 & 9.084 & 8.314 $\pm$ 0.002 & 7.954 $\pm$ 0.003 & 7.735 $\pm$ 0.005 & 6.779 $\pm$ 0.013 & 3.381 $\pm$ 0.033 \\[10pt]
1211 & 16.678 & 14.878 & 15.280 & 13.019 & 10.749 & 9.795 & 9.230 & 8.233 $\pm$ 0.002 & 7.824 $\pm$ 0.003 & 7.677 $\pm$ 0.004 & 7.022 $\pm$ 0.002 & 4.375 $\pm$ 0.046 \\
1212 & 14.380 & 13.489 & 12.558 & 11.974 & 10.765 & 9.844 & 8.963 & 6.685 $\pm$ 0.001 & 5.703 $\pm$ 0.001 & 5.152 $\pm$ 0.001 & 3.974 $\pm$ 0.001 & 0.95 $\pm$ 0.034\\
1247 & 19.881 & 18.179 & 16.168 & 15.021 & 13.060 & 12.324 & 11.891 & 11.209 $\pm$ 0.005 & 10.841 $\pm$ 0.008 & 10.567 $\pm$ 0.014 & 9.827 $\pm$ 0.013 & 7.519 $\pm$ 0.032 \\
1259 & 19.901 & 17.900 & 16.406 & 15.376 & 13.035 & 11.860 & 11.176 & 10.065 $\pm$ 0.004 & 9.537 $\pm$ 0.004 & 9.324 $\pm$ 0.018 & 8.725 $\pm$ 0.066 & 5.800 $\pm$ 0.045 \\
1262 & 16.508 & 15.819 & 14.004 & 13.217 & 11.522 & 10.537 & 9.911 & 8.811 $\pm$ 0.003 & 8.423 $\pm$ 0.004 & 8.135 $\pm$ 0.004 & 7.448 $\pm$ 0.003 & 4.712 $\pm$ 0.037 \\
1284 & 20.059 & 18.223 & 17.161 & 15.850 & 13.759 & 13.089 & 12.671 & 11.737 $\pm$ 0.008 & 11.158 $\pm$ 0.007 & 10.863 $\pm$ 0.019 & 10.168 $\pm$ 0.012 & 7.069 $\pm$ 0.031 \\
1333 & 16.519 & 15.230 & 14.581 & 14.226 & 12.644 & 12.102 & 11.832 & 11.620 $\pm$ 0.007 & 11.603 $\pm$ 0.010 & 11.588 $\pm$ 0.024 & 11.504 $\pm$ 0.030 & \nodata \\
1643 & 19.202 & 17.376 & 15.534 & 14.453 & 12.565 & 11.753 & 11.406 & 11.162 $\pm$ 0.007 & 11.054 $\pm$ 0.008 & 11.034 $\pm$ 0.022 & 10.957 $\pm$ 0.022 & \nodata \\
\enddata
\tablenotetext{a}{The source is not resolved from a nearby star in the J band and only an upper limit is reported.}
\tablecomments{SDSS u band data is not reported here because of the red leak in the filter.}
\end{deluxetable}

\begin{deluxetable}{lcccccccc}
\rotate
\tablewidth{0pt}
\tabletypesize{\small}
\tablecaption{Cluster members identified from Low-Resolution Spectra\label{lowres}}
\tablehead{\colhead{Object ID}&\colhead{RA (J2000)}&\colhead{Dec (J2000)}&\colhead{Spectral Type Range}&\colhead{Adopted Spectral Type}&\colhead{H$\alpha$EW}&\colhead{Li EW}&\colhead{CTTS/WTTS}\\\colhead{}&\colhead{}&\colhead{}&\colhead{}&\colhead{}&\colhead{($\AA$)}&\colhead{($\AA$)}&\colhead{}}
\startdata
17&05 45 21.36&-00 00 45.72&M3&M3&4.4&0.2&WTTS\\
29&05 45 22.68&-00 14 27.13&M6&M6&12.2&0.55&WTTS\\
63&05 45 26.16&00 06 37.91&M3&M3&20.8&0.35&CTTS\\
141&05 45 38.26&-00 08 10.97&K7-M0&M0&4.1&0.43&WTTS\\
174&05 45 41.68&-00 04 02.42&M2-M4&M3&5.4&0.4&WTTS\\
177&05 45 41.94&-00 12 05.33&K3-K4&K4&1.2&0.48&WTTS\\
204\tablenotemark{\dagger}&05 45 44.37&00 22 58.22&M0-M2&M1&29.4&0.6&CTTS\\
229&05 45 46.91&00 14 29.36&M3-M5&M4&6.8&0.7&WTTS\\
281&05 45 53.11&-00 13 24.89&M1&M1&18.1&0.59&CTTS\\
285\tablenotemark{\dagger}&05 45 53.54&00 33 08.82&M1-M2&M2&150&0.15&CTTS\\
288&05 45 53.60&00 22 42.10&K7&K7&2.4&0.28&WTTS\\
322&05 45 56.20&00 33 10.33&M5-M6&M5.5&8.3&0.4&WTTS\\
326&05 45 56.31&00 07 08.58&K7-M0&K9&6.2&0.68&CTTS\\
344&05 45 57.38&00 20 22.20&K6-M0&K7&35.3&0.5&CTTS\\
351\tablenotemark{\dagger}&05 45 57.93&00 02 48.55&M3-M5&M4&75&0.4&CTTS\\
373&05 46 00.18&00 03 07.06&M4&M4&27.2&0.36&CTTS\\
416&05 46 04.58&00 00 38.16&M2&M2&7.4&0.2&WTTS\\
418\tablenotemark{a}\tablenotemark{\dagger}&05 46 04.64&00 04 58.15&K7-M1&K9&126&0.26&CTTS\\
458&05 46 07.89&-00 11 56.87&K2-K3&K3&3.2&0.58&CTTS\\
480&05 46 09.27&00 13 32.63&M0-M1&M1&10.8&0.48&CTTS\\[10pt]
515&05 46 11.86&00 32 25.91&K1-K2&K2&3.8&0.7&CTTS?\\
543&05 46 14.48&00 20 24.36&M3-M4&M4&3.6&$<$0.35&WTTS\\
571&05 46 18.30&00 06 57.85&K1-K3&K1&34.7&0.4&CTTS\\
581&05 46 18.89&-00 05 38.11&K3-K4&K4&3.9&0.73&CTTS\\
584\tablenotemark{\dagger}&05 46 19.06&00 03 29.59&K4-K6&K5&44.5&0.44&CTTS\\
590&05 46 19.47&-00 05 20.00&K2-K3&K2.5&28.5&0.5&CTTS\\
601&05 46 20.88&00 08 09.42&K6-M1&K8&8.7&0.7&CTTS\\
618&05 46 22.44&-00 08 52.62&K1-K2&K1&32&0.55&CTTS\\
628&05 46 22.99&00 04 26.44&M1-M3&M2&104&0.5&CTTS\\
656&05 46 25.89&00 09 31.97&M1-M2&M2&32.4&$<$0.4&CTTS\\
663&05 46 26.65&00 31 07.50&K4-K8&K6&4.9&0.6&WTTS\\
677&05 46 27.83&00 05 48.37&M2-M3&M2.5&3&0.3&WTTS\\
689&05 46 29.00&00 29 07.19&M1-M2&M1.5&4.8&0.5&WTTS\\
699&05 46 30.06&00 12 09.68&M1-M2&M1.5&15.4&0.5&CTTS\\
712\tablenotemark{\dagger}&05 46 31.05&00 25 33.37&continuum&-&92.3&0.25&CTTS\\
730&05 46 33.28&00 02 51.90&K5-M0&K7&24.4&0.6&CTTS\\
739\tablenotemark{b}&05 46 34.54&00 06 43.45&K3-K5&K4&5.5&0.5&CTTS\\
741&05 46 34.90&00 04 20.68&M2-M3&M2.5&5&0.2&WTTS\\
760&05 46 37.06&00 01 21.79&K4-K7&K5.5&61&0.4&CTTS\\
782&05 46 38.57&00 22 05.99&M0-M1&M1&5.8&0.2&WTTS\\
802&05 46 40.18&00 05 01.86&M0-M1&M1&7.8&0.8&WTTS\\[10pt]
813\tablenotemark{\dagger}&05 46 40.77&00 27 22.50&M2-M4&M3&166&0.1&CTTS\\
843&05 46 43.85&00 15 32.36&M2&M2&4.9&0.6&WTTS\\
849\tablenotemark{c}&05 46 44.09&00 18 03.17&K5-K7&K6&7.1&0.7&CTTS\\
856&05 46 44.84&00 16 59.77&M1-M2&M1&14&0.4&CTTS\\
878\tablenotemark{\dagger}&05 46 46.87&00 09 07.63&M0&M0&45&0.4&CTTS\\
934\tablenotemark{c}&05 46 51.48&00 19 21.32&M1-M2&M1.5&25.1&0.3&CTTS\\
941&05 46 52.41&00 20 01.68&K7-M1&K9&41&0.5&CTTS\\
984\tablenotemark{c}&05 46 56.54&00 20 52.91&K2-K4&K3&2.3&0.7&WTTS\\
997\tablenotemark{\dagger}&05 46 58.03&00 14 27.82&M3-M4&M3&19&0.45&CTTS\\
998&05 46 58.13&00 05 38.15&K7-M0&M0&2&0.73&WTTS\\
1062&05 47 03.97&00 11 14.35&M1-M2&M2&42&0.45&CTTS\\
1078\tablenotemark{c}&05 47 04.94&00 18 31.64&M0-M1&M0.5&2.3&0.4&WTTS\\
1089&05 47 05.34&00 28 46.13&M1-M2&M1.5&6&0.3&WTTS\\
1099\tablenotemark{\dagger}&05 47 06.00&00 32 08.48&K0-K3&K0&20&0.7&CTTS\\
1116&05 47 06.96&00 00 47.74&K3-K6&K4.5&28.4&0.5&CTTS\\
1117&05 47 06.99&00 31 55.96&M0-M2&M1&63&0.3&CTTS\\
1119\tablenotemark{c}&05 47 07.26&00 19 32.23&K7-M0&M0&48&0.3&CTTS\\
1134&05 47 08.69&00 00 14.04&M2-M4&M3&9.4&0.42&WTTS\\
1171&05 47 10.89&00 32 05.96&M1-M2&M1.5&8.4&0.57&WTTS\\
1173\tablenotemark{c}&05 47 10.98&00 19 14.81&G4-G8&G6&17&0.6&CTTS\\[10pt]
1211\tablenotemark{\dagger}&05 47 13.85&00 00 17.06&M0-M2&M1&30.4&0.7&CTTS\\
1212\tablenotemark{\dagger}&05 47 14.11&00 09 07.34&continuum&-&78&0.06&CTTS\\
1247&05 47 16.58&-00 00 56.38&M4-M5&M4&10.3&0.36&WTTS\\
1259\tablenotemark{\dagger}&05 47 17.16&00 18 24.59&M0-M2&M1&65.5&0.3&CTTS\\
1262&05 47 17.28&00 38 21.37&K5-M0&K7&8.3&0.6&CTTS\\
1284\tablenotemark{\dagger}&05 47 19.72&00 01 21.83&M5-M6&M5&300&0.47&CTTS\\
1333&05 47 25.05&00 31 04.94&K6-K7&K7&1.3&0.24&WTTS\\
1643&05 47 57.53&00 11 31.24&M3-M5&M4&9.5&0.4&WTTS\\
\enddata
\tablenotetext{\dagger}{[O I] 6300 \AA\ emission detected in optical spectra}
\tablenotetext{a}{Variable in the optical as observed by \citet{bri05}. They measured a spectral type of K6, and \ha\ and Li EWs of 98 and 0.4\AA\ respectively.}
\tablenotetext{b}{Binary from \citet{pad97}.}
\tablenotetext{c}{X-ray emission detected by \citet{ski07}}
\end{deluxetable}

\begin{deluxetable}{lcccccccc}
\tablewidth{0pt}
\tabletypesize{\small}
\tablecaption{Stellar Parameters for Likely Members}
\tablehead{\colhead{Object ID}&\colhead{Luminosity}&\colhead{$T_{eff}$}&\colhead{Mass}&\colhead{Radius}&\colhead{$A_{V}$}&\colhead{Accretion Rate}&\colhead{\alphairac}&\colhead{Disk?}\\\colhead{}&\colhead{$L_{\sun}$}&\colhead{K}&\colhead{$M_{\sun}$}&\colhead{$R_{\sun}$}&\colhead{}&\colhead{$10^{-8}$\accunits}&\colhead{}&\colhead{}}
\startdata
17&0.298&3470&0.35&1.54&1.9&-&-2.65&no\\
29&0.068&3050&0.15&0.95&0.0&-&-2.64&no\\
63&0.348&3470&0.34&1.66&1.1&0.79&-1.28&strong\\
141&0.514&3850&0.57&1.64&2.3&0&-2.70&no\\
174&0.277&3470&0.34&1.48&1.6&-&-2.77&no\\
177&1.051&4590&1.30&1.65&1.8&0&-2.58&Transition\\
204&0.463&3720&0.47&1.67&1.1&0.83&-1.32&strong\\
229&0.077&3370&0.26&0.83&1.2&-&-2.79&no\\
281&0.235&3720&0.47&1.19&1.1&0.24&-2.39&Transition\\
285&0.348&3580&0.39&1.56&1.5&0.87&-1.12&strong\\
288&0.230&4060&0.79&0.99&0.7&0&-2.71&no\\
322&0.214&3240&0.25&1.49&1.5&-&-2.61&no\\
326&0.687&3990&0.68&1.77&1.6&$<$0.47&-1.79&IRAC-weak\\
344&0.745&3990&0.69&1.84&1.6&1.79&-1.52&strong\\
351&0.091&3370&0.28&0.90&0.8&-&-1.76&IRAC-weak\\
373&0.496&3370&0.30&2.10&1.5&1.65&-1.15&strong\\
416&0.319&3580&0.39&1.49&3.1&-&-1.45&strong\\
418&0.895&3990&0.67&2.01&2.2&2.27&-1.03&strong\\
458\tablenotemark{a}&4.679&4730&2.07&3.33&0.0\tablenotemark{b}&$<$7.11\tablenotemark{c}&-1.91&IRAC-weak\\
480&0.807&3720&0.47&2.20&1.4&1.77&-1.13&strong\\[10pt]
515&2.280&4900&1.55&2.13&1.3&$<$1.09&-2.23&IRAC-weak\\
543&0.116&3370&0.30&1.01&1.8&-&-1.92&IRAC-weak\\
571&2.407&4900&1.65&2.19&4.7&$<$1.10&-0.75&strong\\
581&3.576&4590&1.56&3.04&3.8&$<$2.37&-1.25&MIPS-weak\\
584&3.280&4350&1.12&3.244&4.5&7.83&-1.06&strong\\
590&3.525&4815&1.85&2.74&2.6&3.22&-0.72&strong\\
601&0.963&3955&0.64&2.13&4.3&-&-0.75&strong\\
618&2.053&5080&1.26&1.88&2.6&$<$1.06&-1.67&IRAC-weak\\
628&0.391&3580&0.39&1.65&3.4&-&-1.02&strong\\
656&0.352&3580&0.39&1.57&3.1&-&-1.24&strong\\
663&1.419&4205&0.89&2.28&4.4&-&-1.27&MIPS-weak\\
677&0.643&3525&0.37&2.19&4.1&-&-2.64&no\\
689&0.721&3650&0.43&2.16&2.2&0&-2.66&no\\
699&0.899&3650&0.43&2.41&4.6&-&-1.21&strong\\
730&1.280&4025&0.72&2.37&3.9&-&-1.06&strong\\
739&3.182&4590&1.63&2.87&6.2&-&-1.16&MIPS-weak\\
741&0.625&3525&0.37&2.16&4.0&-&-1.82&IRAC-weak\\
760&1.655&4277&1.01&2.38&3.5&5.13&-1.13&strong\\
782&0.491&3720&0.47&1.72&4.0&-&-1.39&MIPS-weak\\
802&0.552&3720&0.47&1.82&2.7&-&-2.19&IRAC-weak?\\[10pt]
813&0.335&3470&0.34&1.63&3.2&-&-1.28&strong\\
843&0.399&3580&0.39&1.67&2.4&-&-1.56&strong\\
849&2.006&4205&0.90&2.71&2.9&$<$1.79&-1.26&MIPS-weak\\
856&0.618&3720&0.47&1.93&4.2&-&-1.83&IRAC-weak\\
878&0.605&3850&0.57&1.78&3.2&1.09&-0.73&strong\\
934&0.422&3650&0.42&1.65&3.2&-&-1.09&strong\\
941&0.505&3920&0.62&1.57&2.8&1.54&-0.96&strong\\
984&2.094&4730&1.89&2.19&4.5&$<$1.78&-2.44&IRAC-weak?\\
997&0.315&3470&0.34&1.58&3.4&-&-0.93&strong\\
998&3.269&3850&0.56&4.13&3.7&$<$5.44&-1.79&IRAC-weak\\
1062&0.441&3580&0.39&1.76&2.0&1.94&-0.91&strong\\
1078&1.973&3785&0.51&3.32&4.2&-&-2.24&IRAC-weak?\\
1089&0.299&3650&0.42&1.39&3.4&-&-2.64&no\\
1099&2.361&4900&1.64&2.17&5.5&-&-1.11&strong\\
1116&1.532&4470&1.29&2.10&1.8&$<$0.81&-0.97&strong\\
1117&0.362&3720&0.47&1.47&3.8&-&-0.56&strong\\
1119&7.294&3850&0.56&6.18&3.8&$<$18.2&-0.80&MIPS-weak\\
1134&0.667&3470&0.35&2.30&1.2&0&-2.62&no\\
1171&0.519&3720&0.47&1.74&2.4&-&-1.28&MIPS-weak\\
1173&9.082&5700&1.84&3.14&5.1&no stand&-1.13&strong\\[10pt]
1211&1.302&3720&0.47&2.80&1.6&7.75&-1.51&strong\\
1247&0.253&3370&0.30&1.50&2.5&-&-1.29&MIPS-weak\\
1259&0.336&3720&0.47&1.42&3.3&-&-1.37&strong\\
1262&2.798&4025&0.70&3.50&4.5&8.60&-1.31&strong\\
1284&0.108&3240&0.22&1.06&1.4&-&-1.10&strong\\
1333&0.216&4060&0.79&0.96&0.5&0&-2.71&no\\
1643&0.348&3370&0.30&1.76&1.9&-&-2.62&no\\
\enddata
\tablenotetext{a}{Photometry may be contaminated by background source with very cool SED, possible a Class 0/I source.}
\tablenotetext{b}{With no available optical photometry, and extended emission contaminating the 2MASS photometry, an accurate extinction could not be derived and we assume a lower limit of 0}
\tablenotetext{c}{R band magnitude from \citet{s00} 2 Myr isochrone used instead of observed R band magnitude to derive accretion rate from veiling.}
\tablecomments{A dash in the accretion rate column indicates no high-resolution spectra was available, a 0 is used for those stars with high-resolution spectra, but the \ha\ profile indicated there was no active accretion. Disk descriptors: strong (\alphairac$>-1.6$, [8.0]-[24]$>2.4$), MIPS-weak (\alphairac$<-1.6$, [8.0]-[24]$<2.4$), IRAC-weak($-1.6>$\alphairac$>-2.5$), gap (small short wanvelength excess, large [24] excess), no disk (\alphairac$<-2.5$).\label{member_params}}
\end{deluxetable}

\begin{deluxetable}{lcccccc}
\tablewidth{0pt}
\tablecaption{High-Resolution Spectra of Likely Members}
\tablehead{\colhead{Object ID}&\colhead{Radial Velocity}&\colhead{Vsini}&\colhead{T\&D R\tablenotemark{a}}&\colhead{\ha\ FW 10\%}&\colhead{\ha EW}&\colhead{Veiling}\\\colhead{}&\colhead{(km/sec)}&\colhead{(km/sec)}&\colhead{}&\colhead{(km/sec)}&\colhead{$\AA$}&\colhead{}}
\startdata
63&31.7$\pm$3.0&$<$8&3.7&280&19.1&0.88\\
141&27.0$\pm$1.3&$<$8&12.0&142&1.7&-\\
177&27.7$\pm$0.8&$<$8&23.6&124(101)&0.9(1.2)&-\\
204&25.5$\pm$1.2&$<$8&11.5&250&25.4&0.51\\
281&27.3$\pm$1.3&$<$8&16.7&400&32.4&0.36\\
285&26.2$\pm$3.6&14.8&3.9&600&120.7&0.94\\
288&30.2$\pm$1.3&$<$8&13.4&110&1.4&-\\
326\tablenotemark{b}&27.0$\pm$1.1&$<$8&15.5&170(170)&5.9(5.3)&-\\
344&21.6$\pm$1.3&$<$8&12.1&495&27.4&0.76\\
373&-&-&-&290&24.8&2.2\\
418&27.0$\pm$2.1&13.4&7.4&410&60.5&0.76\\
458&24.5$\pm$1.1&29.7&18.4&280(320)&2.7(3.4)&$<$0.15\\
480&26.8$\pm$1.3&$<$8&12.5&510&7.9&0.45\\
515\tablenotemark{b}&14.1$\pm$1.0&24.3&20.1&410&3.4&$<$0.15\\
571&28.8$\pm$1.5&15.7&11.3&483&29.6&$<$0.15\\
581&25.9$\pm$0.9&18.3&21.2&480&4.8&$<$0.15\\
584&-&-&-&640&42.7&0.39\\
590&28.2$\pm$0.7&$<$8&23.1&470&14.9&0.31\\
618&27.0$\pm$0.7&26.8&19.2&530(520)&46.8(28.3)&$<$0.15\\
689&27.2$\pm$1.2&$<$8&10.9&110&3.7&-\\[10pt]
760&28.5$\pm$1.8&$<$8&8.2&535&49.4&1.28\\
849&27.3$\pm$1.6&$<$8&9.8&340&18.2&$<$0.15\\
878&27.1$\pm$1.3&$<$8&12.6&440&25.5&0.49\\
941&26.8$\pm$2.3&$<$8&6.5&415&52.4&1.82\\
984&25.4$\pm$2.4&49.2&10.6&280&1.1&$<$0.15\\
998&11.6$\pm$2.3&40.7&9.7&280&1.9&$<$0.15\\
1062&26.9$\pm$2.9&$<$8&4.3&430&46.8&3.18\\
1116&25.8$\pm$0.8&$<$8&21.9&400(405)&23.4(22.1)&$<$0.15\\
1119&31.3$\pm$3.8&55.8&2.9&450&27.1&$<$0.15\\
1134&27.1$\pm$1.5&$<$8&8.7&160&8.0&-\\
1173&32.7$\pm$2.3&49.1&11.4&350&9.7&-\\
1211&25.9$\pm$2.5&$<$8&5.2&490&26.3&1.97\\
1262&26.2$\pm$1.2&$<$8&14.1&340&5.4&0.41\\
1333&28.3$\pm$1.1&$<$8&15.6&90&1.3&-\\
\enddata
\tablecomments{FW at 10\% values in parenthesis are from the second set of high-resolution spectra. Only objects 326,177,458,618 and 1116 have two sets of high-resolution spectra.\label{highres}}
\tablenotetext{a}{Tonry \& Davis R value \citep{td79}. The 90\% confidence interval for v sini is approximately vsini/(1+R).}
\tablenotetext{b}{\ha\ profile shows an inverse P Cygni shape indicating active accretion, despite the small FW 10\%.}
\end{deluxetable}

\begin{deluxetable}{ccccc}
\tabletypesize{\footnotesize}
\tablewidth{0pt}
\tablecaption{Fraction of Evolved Disks\label{evolved_fraction}}
\tablehead{\colhead{Cluster}&\colhead{Age}&\colhead{Strong disks}&\colhead{MIPS-weak}&\colhead{IRAC-weak}}
\startdata
\cutinhead{K0-M1}
Taurus&1&$94\pm24\%\ (16/17)$&-&$6\pm6\%\ (1/17)$\\
NGC2068/71&2&$66\pm14\%\ (23/35)$&$16\pm7\%\ (5/32)$&$20\pm8\%\ (7/35)$\\
IC 348&2-3&$80\pm20\%\ (16/20)$&$0\%\ (0/17)$&$20\pm10\%\ (4/20)$\\
$\sigma$ Ori&3&$75\pm19\%\ (15/20)$&$19\pm11\%\ (3/16)$&$10\pm7\%\ (2/20)$\\
Tr 37&4&$68\pm9\%\ (59/87)$&$13\pm5\%\ (7/54)$&$24\pm5\%\ (21/87)$\\
NGC 2362&5&$30\pm17\%\ (3/10)$&-&$70\pm26\%\ (7/10)$\\
\enddata
\tablecomments{Statistics for MIPS-weak disks only include those stars detected at 24\micron. IRAC-weak disks defined as $-2.56<$\alphairac$<-1.8$.}
\end{deluxetable}

\end{document}